\documentclass[useAMS]{mn2e}
\pdfoutput=1
\usepackage{epsfig}
\usepackage{amsmath}
\usepackage{amssymb}
\usepackage{appendix}
\def\be{\begin{equation}}
\def\ee{\end{equation}}
\def\ba{\begin{eqnarray}}
\def\ea{\end{eqnarray}}
\def\go{\mathrel{\raise.3ex\hbox{$>$}\mkern-14mu
             \lower0.6ex\hbox{$\sim$}}}
\def\lo{\mathrel{\raise.3ex\hbox{$<$}\mkern-14mu
             \lower0.6ex\hbox{$\sim$}}}
\def\bxi{{\mbox{\boldmath $\xi$}}}
\def\br{{\bf r}}

\def\bomega{{\bar\omega_\alpha}}

\begin{document}

\title[Tidal Synchronization and Dissipation in White Dwarf Binaries]
{Dynamical Tides in Compact White Dwarf Binaries: Tidal Synchronization and Dissipation}
\author[J. Fuller and D. Lai]
{Jim Fuller\thanks{Email:
derg@astro.cornell.edu; dong@astro.cornell.edu}
and Dong Lai\\
Center for Space Research, Department of Astronomy, Cornell University, Ithaca, NY 14853, USA}

\label{firstpage}
\maketitle

\begin{abstract}
In compact white dwarf (WD) binary systems (with periods ranging from
minutes to hours), dynamical tides involving the excitation and
dissipation of gravity waves play a dominant role in determining the
physical conditions (such as rotation rate and temperature) of the WDs
prior to mass transfer or binary merger.  We calculate the amplitude
of the tidally excited gravity waves as a function of the tidal
forcing frequency $\omega=2(\Omega-\Omega_s)$ (where $\Omega$ is the
orbital frequency and $\Omega_s$ is the spin frequency) for several realistic
carbon-oxygen WD models, under the assumption that the outgoing
propagating waves are efficiently dissipated in outer layer of the
star by nonlinear effects or radiative damping. Unlike main-sequence
stars with distinct radiative and convection zones, the mechanism of
wave excitation in WDs is more complex due to the sharp features
associated with composition changes inside the WD.  In our WD models,
the gravity waves are launched just below the helium-carbon boundary
and propagate outwards.  We find that the tidal torque on the WD and
the related tidal energy transfer rate, $\dot E_{\rm tide}$, depend on
$\omega$ in an erratic way, with $\dot E_{\rm tide}$ varying
by orders of magnitude over small frequency ranges. On average,
$\dot E_{\rm tide}$ scales approximately as $\Omega^5\omega^5$ for a
large range of tidal frequencies.

We also study the effects of dynamical tides on the long-term
evolution of WD binaries prior to mass transfer or merger. Above
a critical orbital frequency $\Omega_c$, corresponding to an orbital period
of order one hours(depending on WD models), dynamical tides efficiently 
drive $\Omega_s$ toward $\Omega$, although a small, almost constant
degree of asynchronization ($\Omega-\Omega_s\sim {\rm constant}$) is maintained
even at the smallest binary periods. While the orbital decay is always
dominated by gravitational radiation, the tidal energy transfer 
can induce significant phase error in the low-frequency 
gravitational waveforms, detectable by the planned LISA project.
Tidal dissipation may also lead to significant heating of the WD envelope
and brightening of the system long before binary merger.

\end{abstract}


\section{Introduction}

Compact white dwarf (WD) binary systems (with orbital periods in the
range of minutes to hours) harbor many interesting and unanswered
astrophysical questions. An increasing number of such systems are
being discovered by recent surveys (e.g. Mullally et al. 2009;
Kulkarni \& van Kerkwijk 2010; Steinfadt et al. 2010; Kilic et
al. 2011; Brown et al. 2011; see Marsh 2011 for a review). The orbits of these systems
decay via the emission of gravitational waves, which could be detected by the 
planned {\it Laser Interferometer Space Antenna (LISA)} (Nelemans
2009). Depending on the WD masses and the physics of the merger process,
these merging WD systems may produce single helium-rich sdO stars,
giant stars (R CrB stars), stable mass transfer AM CVn binaries, or
possibly underluminous supernovae. Most importantly, compact WD binaries in which the total mass is near
the Chandrasekhar limit are thought to be the probable
progenitors of type Ia supernovae upon a stellar merger at the end of
the orbital decay process (Webbink 1984; Iben \& Tutukov 1984). Recent
studies have provided support for this \textquotedblleft double
degenerate\textquotedblright \ scenario (e.g., Gilfanov \& Bogdan 2010;
Di Stefano 2010; Maoz et al. 2010)
and even sub-Chandrasekhar WD mergers may lead to type
Ia supernovae (van Kerkwijk et al. 2010).

Prior to merger, tidal interactions may affect the properties of the
binary WDs and their evolutions, including the phase evolution
of the gravitational waves. Previous studies have focused on
equilibrium tides (e.g., Iben et al. 1998; Willems et al. 2010),
corresponding to quasi-static deformation of the star. Such
equilibrium tides are unlikely to play a role in the tidal
synchronization/dissipation process.
Iben et al. (1998) estimated the
effect of tidal heating in the WD based on the assumption that the
(spherically averaged) local heating rate is equal to the rate of
rotational energy deposition required to maintain
synchronization. They suggested that the binary WDs may brighten by
several magnitudes before merger. 

In fact, in a compact WD binary, as the orbital decay rate due to
gravitational wave radiation increases rapidly with decreasing orbital
period, it is not clear if tidal effects are sufficiently strong to
drive the binary system toward synchronous rotation. The critical
orbital period for synchronization is unknown. For this reason, the
majority of recent WD merger simulations (e.g., Segretain et al. 1997;
Loren-Aguilar et al. 2009; Pakmor et al.~2010,2011)
have assumed the merging WDs to be
non-synchronized prior to merger. However, whether the WDs are spin-synchronized
may affect the merger product and the possible supernova signature:
for example, the strong velocity shear between the stars upon contact
would be significantly reduced for the merger of a synchronized binary.
The degree of synchronization also determines the tidal luminosity 
of the binary prior to merger. Indeed, it is possible that tidal dissipation contributes significantly to the brightness of some of the recently observed WD binaries (e.g., Brown et al. 2011).

In a recent paper (Fuller \& Lai 2011, hereafter Paper I), we used
linear theory to calculate the the excitation of discrete gravity
modes in a WD due to the tidal gravitational field of a compact
companion star (a WD, neutron star or black hole). The existence of
discrete modes requires that gravity waves be reflected near the
surface of the WD. In this case, tidal energy and angular momentum
transfers between the WD and the binary orbit occur only during a
series of resonances, when the g-mode frequency $\sigma_\alpha$ equals
$2\Omega$ (where $\Omega$ is the orbital frequency).
Our calculations showed that while the dimensionless (mass-weighted)
amplitude of the resonantly excited g-mode is not extremely non-linear 
(it approaches $\approx 0.1$), the displacement associated with the mode 
becomes large in the outer layer of the WD where the density is low. 
In other words, while the mode does
not reach a non-linear amplitude in the bulk interior of the star, it becomes
very non-linear in the outer layers even before resonance. 
We concluded that tidally excited gravity waves are likely to continually 
damp in the outer layer of the WD, preventing the formation of discrete modes.
A proper treatment of dynamical tides in binary WDs must take account of
this continuous wave damping. 

In this paper, we calculate the tidal excitation of gravity waves
in binary WDs assuming that the waves are efficiently damped
in the WD envelope. To this end, we implement an outgoing wave 
boundary condition near the WD surface. 
Unlike gravity modes (which have a set of discrete eigenfrequencies), the outgoing wave
boundary condition permits the excitation of gravity waves at all
frequencies, and thus allows for a continuous process of tidal
dissipation. Similar calculations have been implemented for 
early-type stars (Zahn 1975,1977; Goldreich \& Nicholson 1989) and
solar-type stars (Goodman \& Dickson 1998; Ogilvie \& Lin 2007).
In early-type stars, gravity waves are excited at the boundary between
the convective core and radiative envelope, propagate outwards and dissipate
in the outer envelope. In solar-type stars, gravity waves are similarly 
excited at the interface between the radiative core and convective envelope,
but propagate inward before dissipating (via non-linear wave
breaking) near the center of the star (Barker \& Ogilvie 2010,2011).
Unlike main-sequence stars, WDs do not contain a simple two-zone
structure of convective and radiative regions, and it is not clear how
and to what extent gravity waves are excited.
The outgoing wave outer boundary condition allows us to calculate the
rate at which energy and angular momentum are transferred to the WD as
a function of orbital period. We can then calculate the orbital period
at which tidal effects can compete with orbital decay due to
gravitational radiation. At this orbital period, the synchronization
process can begin. Furthermore, by scaling our results to rotating
WDs, we can determine the WD spin period and energy dissipation rate
at any orbital period. 

This paper is organized as follows. In Sections 2-4 we derive the
basic equations for tidally forced stellar oscillations, the boundary
conditions and the tidal angular momentum and energy transfer rates.
In Section 5 we discuss our numerical method and present several test
calculations, where we emphasize the importance of using a 
self-consistent stellar model in order to obtain reliable amplitudes
for tidally excited gravity waves.  In Section 6 we present our
numerical calculations of tidal excitations for realistic WD
models. Since our numerical results reveal a complicated dependence of
the tidal energy transfer rate on the tidal frequency, we examine
a simple semi-analytic model in Section 7 to shed light on the mechanism 
of gravity wave excitation. In Section 8 we use the results of previous sections to 
study the long-term spin-orbit evolution of WD binaries,
including spin synchronization, the tidal effect on the low-frequency
gravitational radiation waveforms and tidal heating of the WDs.
We conclude in Section 9 with a discussion of theoretical uncertainties and future work.

%%%%%%%%%%%%%%%%%%%%%%%%%%%%%%%%%%%%%%%%%%%%%%%%%%%%%%%%%%
\section{Basic Equations}

The dynamical tide of the WD (mass $M$) is driven by the external gravitational potential of the the companion (mass $M'$). The leading order (quadrupole) potential is 
\be
\label{U}
U_{\rm ex}({\bf r},t) = U(r) \Big[ Y_{22}(\theta,\phi)e^{-i \omega t} +  Y_{22}^*(\theta,\phi) e^{i \omega t} \Big]
\ee
with  
\be
\label{U2}
U(r) = - \frac{G M' W_{22}}{a^3} r^2 .
\ee
Here $a$ is the orbital separation, $\omega = 2 \Omega$ is the tidal frequency for a non-spinning WD (we will account for the spin effect in Section \ref{spin}), $\Omega$ is the orbital frequency, and $W_{22} = \sqrt{3 \pi/10}$. The actual fluid perturbations in the WD can be written as $\bxi_{\rm ac}({\bf r},t) = \bxi({\bf r},t) + \bxi^*(\br,t)$ for the Lagrangian displacement and $\delta P_{\rm ac}(\br,t) = \delta P({\bf r},t) + \delta P^*(\br,t)$ for the Eulerian pressure perturbation, and similarly for other quantities. In the following, we shall consider perturbations ($\bxi$, $\delta P$, etc.) driven by the potential $U({\bf r},t) = U(r) Y_{22}(\theta,\phi) e^{-i \omega t}$. We shall adopt the Cowling approximation (so that the gravitational potential perturbation $\delta \Phi$ associated with the density perturbation is neglected, i.e., $\delta \Phi =0$), which is valid for gravity waves in the star. We will consider adiabatic oscillations, for which the Lagrangian perturbations in pressure and density are related by $\Delta P = a_s^2 \Delta \rho$, where $a_s$ is the adiabatic sound speed. This is a good approximation in the bulk of the star where the thermal time is much longer than the wave period.

Letting 
\be
\label{Pib}
\delta P({\bf r},t) = \delta P (r) Y_{22}(\theta,\phi) e^{-i \omega t},
\ee
and
\be
\label{xib}
\bxi({\bf r},t) = \big[\xi_r(r) \hat{r} + \xi_\perp (r) r \nabla_\perp \big] Y_{22}(\theta,\phi) e^{-i \omega t},
\ee
the fluid perturbation equations reduce to  
\be
\label{xir'}
\frac{1}{r^2} \big(r^2 \xi_r \big)' - \frac{g}{a_s^2} \xi_r + \frac{1}{\rho a_s^2} \bigg(1 - \frac{L_l^2}{\omega^2} \bigg) \delta P - \frac{l(l+1)U}{\omega^2 r^2} = 0,
\ee
and
\be
\label{p'}
\frac{1}{\rho} \delta P' + \frac{g}{\rho a_s^2}\delta P + \big(N^2 - \omega^2\big)\xi_r + U' = 0,
\ee
where the $'$ denotes $d/dr$. In equations (\ref{xir'}) and (\ref{p'}), $L_l$ and $N$ are the Lamb and Br\"unt-Vais\"al\"a frequencies, respectively, given by [note we will continue to use the notations $L_l$, $l(l+1)$, and $m$, although we focus on $l=m=2$ in this paper]
\be 
L_l^2 = \frac{l(l+1)a_s^2}{r^2}
\ee 
and
\be
\label{brunt}
N^2 = g^2 \bigg( \frac{d \rho}{d P} - \frac{1}{a_s^2} \bigg).
\ee
The other perturbation variables are related to $\delta P$ and $\xi_r$ by 
\be
\label{xiperp}
\xi_\perp = \frac{1}{r \omega^2} \bigg( \frac{\delta P}{\rho} + U \bigg),
\ee
\be
\delta \rho = \frac{1}{a_s^2} \delta P + \frac{\rho N^2}{g} \xi_r.
\ee

Defining $Z = \chi^{-1/2} r^2 \xi_r$, where
\be
\chi=\frac{r^2}{\rho a_s^2}\bigg(\frac{L_l^2}{\omega^2} - 1 \bigg),
\ee
equations (\ref{xir'}) and (\ref{p'}) can be combined to yield
\be
\label{z}
Z'' + k^2(r) Z = V(r).
\ee
Here,
\begin{align}
\label{k}
k^2(r) = & \frac{\chi \rho (N^2 - \omega^2)}{r^2} + \frac{1}{2}\bigg(\frac{\chi'}{\chi}\bigg)' - \frac{1}{4} \bigg(\frac{\chi'}{\chi}\bigg)^2 \nonumber \\
& + \frac{g}{a_s^2}\bigg[\frac{-(g/a_s^2)'}{g/a_s^2} + \frac{\chi '}{\chi} - \frac{g}{a_s^2} \bigg]
\end{align}
and 
\be
\label{V}
V(r) = \chi^{-1/2}\bigg[\frac{l(l+1)}{\omega^2}\bigg(\frac{-\chi'}{\chi} + \frac{g}{a_s^2} \bigg) + \frac{2r}{a_s^2} \bigg]U.
\ee

In the WKB limit $|k| \gg 1/H$ and $|k| \gg 1/r$, where $H = |P/P'| \simeq a_s^2/g$ is the pressure scale height, equation (\ref{k}) simplifies to 
\be
\label{kwkb}
k^2(r) = \frac{1}{a_s^2 \omega^2}(L_l^2 - \omega^2)(N^2 - \omega^2). 
\ee
This is the standard WKB dispersion relation for non-radial stellar oscillations (e.g., Unno et al. 1989). For $\omega^2 \ll L_l^2$ and $\omega^2 \ll N^2$, the wave equation (\ref{z}) reduces to
\be
\label{zapproxeq}
Z'' + \frac{l(l+1)(N^2 - \omega^2)}{r^2 \omega^2}Z \simeq -\chi^{-1/2} \frac{l(l+1)N^2}{\omega^2}\frac{U}{g}.
\ee
Then, as long as $|Z''/Z| \gg |\chi''/\chi|$ (which we expect to be true because $Z'' \approx -k^2 Z$ and $\chi'' \sim \chi/H^2$), equation (\ref{zapproxeq}) is identical to the oscillation equations used by Zahn (1975) and Goodman \& Dickson (1998).

\section{Boundary Conditions}

Equations (\ref{xir'}) and (\ref{p'}) or equation (\ref{z}) can be solved with the appropriate boundary conditions at $r=r_{\rm out}$ near the stellar surface and at $r= r_{\rm in} \rightarrow 0$ at the center of the star. The general solution of equation (\ref{z}) can be written as 
\be
Z(r) = c_+ Z_+(r) + c_-Z_-(r) + Z^{\rm eq}(r),
\ee
where $c_+$, $c_-$ are constants. $Z_+(r)$ and $Z_-(r)$ are two independent solutions of the homogeneous equation $Z'' + k^2 Z = 0$, and $Z^{\rm eq}(r)$ represents a particular solution of equation (\ref{z}). We choose the outer boundary $r_{\rm out}$ to be in the wave zone ($k^2 > 0$). If $k^2(r)$ varies slowly such that $|k'/k| \ll k>0$, then the two independent WKB solutions to the homogeneous equation are
\be
Z_{\pm}(r) = \frac{1}{\sqrt{k}} \exp\bigg(\pm i \int^r_{r_o} k dr\bigg), 
\ee
where $r_o$ is an interior point ($r_o < r_{\rm out}$). For $\omega^2 \ll L_l^2$, the WKB wave dispersion relation [equation (\ref{kwkb})] reduces to $\omega^2 = N^2 k_\perp^2/(k^2 + k_\perp^2)$, where $k_\perp^2 = l(l+1)/r^2$, which implies that the radial component of the group velocity is $-\omega k/(k^2 + k_\perp^2)$. Thus, with $\omega > 0$ and $k>0$, $Z_- \propto e^{ -i \int^r_{r_o} k dr}$ represents an outgoing wave, while $Z_+ \propto e^{ i \int^r_{r_o} k dr}$ represents an ingoing wave. An approximate particular solution of equation (\ref{z}) is
\be
\label{zp}
Z^{\rm eq}(r) \simeq \frac{V}{k^2} - \frac{1}{k^2} \bigg(\frac{V}{k^2}\bigg)'',
\ee
where the second term is smaller than the first by a factor of $(kH)^2$ or $(kr)^2$. This represents the \textquotedblleft non-wave\textquotedblright \ equilibrium solution.\footnote{Note that the equilibrium tide usually refers to the f-mode response of the star to the tidal force. Here, we use the term \textquotedblleft equilibrium\textquotedblright \ to refer to the \textquotedblleft non-wave\textquotedblright \ solution.}

Throughout this paper, we adopt the radiative condition at the outer boundary ($r=r_{\rm out}$), i.e., we require that only an outgoing wave exists:
\be
Z(r) \simeq Z^{\rm eq}(r) + \frac{c_-}{\sqrt{k}} \exp \bigg(-i \int^r_{r_o} k dr \bigg).
\ee
This implicitly assumes that waves propagating toward the WD surface are completely damped by radiative diffusion (Zahn 1975) or by non-linear processes. We will check this assumption a posteriori from our numerical results (see Section \ref{nonlinear}). Thus, near the outer boundary, the radial displacement $\xi_r$ behaves as (for $\omega^2 \ll L_l^2$)
\begin{align}
\label{xirout}
\xi_r(r) & = \frac{\chi^{1/2}}{r^2} Z(r) \nonumber \\
& = \xi_r^{\rm eq}(r) + \frac{c_-}{\sqrt{\rho r^2(N^2-\omega^2)}} \exp \bigg(-i \int^r_{r_o} k dr\bigg).
\end{align}
Here $\xi_r^{\rm eq}$ represents the equilibrium tide 
\be
\xi_r^{\rm eq} \simeq \bigg(\!\!-\frac{U}{g}\bigg) \frac{N^2}{N^2-\omega^2} \bigg[1 - \frac{2 g r}{l(l+1) a_s^2} \frac{\omega^2}{N^2}\bigg],
\ee
where we have retained only the first term of equation (\ref{zp}). For $\omega^2 \ll N^2$, this further simplifies to $\xi_r^{\rm eq} \simeq -U/g$ (Zahn 1975). The constant $c_-$ specifies the amplitude of the outgoing wave which is eventually dissipated in the stellar envelope; this is the constant we wish to determine from numerical calculations. 

In practice, to implement equation (\ref{xirout}) at the outer boundary, we require a very accurate calculation of the non-wave solution $\xi_r^{\rm eq}$. This can become problematic when the conditions $|k| \gg 1/H$ and $\omega \ll L_l$ are not well satisfied. Since the transverse displacement $\xi_\perp$ for gravity waves is much larger than the radial displacement in the wave zone, it is more convenient to use $\xi_\perp$ in our outer boundary condition. We define 
\be
Z_1(r) = \bigg(\frac{\rho}{D}\bigg)^{1/2} r^2 \omega^2 \xi_\perp(r),
\ee
with $D \equiv N^2- \omega^2$. Equations (\ref{xir'}) and (\ref{p'}) can be rearranged to yield
\be
Z_1'' + k_1^2(r) Z_1 = V_1 (r),
\ee
where 
\begin{align}
\label{k2}
k_1^2(r) = &-\frac{1}{4} \bigg[\bigg(\ln \frac{\rho r^2}{D} \bigg)'\bigg]^2 - \frac{1}{2}\bigg(\ln\frac{\rho r^2}{D}\bigg)'' - \bigg(\frac{N^2}{g}\bigg)' \nonumber \\
& - \frac{N^2}{g}\bigg(\ln\frac{r^2}{D}\bigg)' + \frac{\omega^2}{a_s^2} + \frac{l(l+1)D}{r^2 \omega^2}
\end{align}
and
\begin{align}
\label{Vo}
V_1(r) = & - \bigg(\frac{\rho r^2}{D}\bigg)^{1/2} \times \nonumber \\
& \Bigg\{ \frac{N^2}{g}U \bigg[\ln \bigg(\frac{r^2 N^2}{D g} U\bigg)\bigg]' - \frac{\omega^2}{a_s^2} U \Bigg\}.
\end{align}
For $k_1 \gg 1/H$ and $\omega^2 \ll L_l^2$, the functions $k_1^2(r)$ and $V_1(r)$ simplify to 
\be
\label{k2approx}
k_1^2(r) \simeq \frac{l(l+1)(N^2-\omega^2)}{\omega^2 r^2}
\ee
and
\be
\label{V1approx}
V_1(r) \simeq - \bigg(\frac{\rho r^2}{D}\bigg)^{1/2} \frac{D}{r^2} \bigg(\frac{Ur^2}{g}\frac{N^2}{D}\bigg)'.
\ee
Again, adopting the radiative boundary condition at $r=r_{\rm out}$, we have
\be
Z_1(r) \simeq \frac{V_1}{k_1^2} - \frac{1}{k_1^2}\bigg(\frac{V_1}{k_1^2}\bigg)'' + \frac{c_-}{k_1} \exp \bigg(-i \int^r_{r_o} k_1 dr\bigg),
\ee
where $c_-$ is a constant. Thus, the transverse displacement $\xi_\perp(r)$ behaves as 
\begin{align}
\label{xiperpout}
\xi_\perp(r) & = \bigg(\frac{D}{\rho}\bigg)^{1/2} \frac{1}{r^2 \omega^2} Z_1(r) \nonumber \\
& = \xi_\perp^{\rm eq} + c_- \bigg(\frac{k_1}{\rho r^2}\bigg)^{1/2} \exp \bigg(-i \int^r_{r_o} k_1 dr\bigg),
\end{align}
where several constants have been absorbed into $c_-$. The equilibrium tidal transverse displacement $\xi_\perp^{\rm eq}(r)$ is given by
\be
\xi_\perp^{\rm eq}(r) \simeq - \frac{1}{l(l+1)r} \bigg(\frac{U r^2}{g}\frac{N^2}{D}\bigg)'.
\ee
For $\omega^2 \ll N^2$, this reduces to (for $l=2$) 
\be
\xi_\perp^{\rm eq}(r) \simeq - \frac{1}{6r} \bigg(\frac{U r^2}{g}\bigg)',
\ee
in agreement with Goldreich \& Nicholson (1989). Thus, we implement the radiative boundary condition at $r=r_{\rm out}$ as 
\be
\label{xiperpoutbc}
\bigg(\xi_\perp - \xi_\perp^{\rm eq}\bigg)' = \Bigg[\frac{-\Big(\rho r^2/k_1\Big)'}{2\Big(\rho r^2/k_1\Big)} - i k_1 \Bigg] \big(\xi_\perp - \xi_\perp^{\rm eq}\big),
\ee
with $\xi_\perp$ computed from $\xi_r$ and $\delta P$ using equation (\ref{xiperp}).

The inner boundary condition can be found by requiring the radial displacement to be finite at the center of the star. This requires
\be
\label{xirinbc}
\xi_r = \frac{l}{\omega^2 r} \bigg( \frac{\delta P}{\rho} + U \bigg) \qquad \big(\textrm{Near} \ r=0\big).
\ee

\section{Angular Momentum and Energy Flux}
\label{ang}

As the wave propagates through the star, it carries an angular momentum flux to the outer layers. At any radius within the star, the $z$ component of the time-averaged angular momentum flux is
\be
\label{Lz}
\dot{J}_z(r) =  \bigg{\langle} \oint d \Omega r^2 \rho \big(\delta v_r + \delta v_r^* \big) \big(\delta v_\phi + \delta v_\phi^* \big) r \sin \theta \bigg{\rangle},
\ee
where $\big{\langle}...\big{\rangle}$ implies time averaging. With
\be
\label{vr}
\delta v_r = -i \omega \xi_r(r) Y_{lm} e^{- i \omega t}
\ee
and
\be
\label{vphi}
\delta v_\phi = -i \omega \xi_\perp(r) r \nabla_\phi Y_{lm} e^{ - i \omega t} = \frac{m \omega \xi_\perp(r)}{\sin \theta} Y_{lm} e^{ - i \omega t},
\ee
we find
\begin{align}
\label{Lz2}
\dot{J}_z(r) & = 2 \oint d \Omega r^3 \rho \omega^2 {\rm Re} \Big[i \xi_r^*(r) Y_{lm}^* m \xi_{\perp}(r) Y_{lm} \Big] \nonumber \\
& = 2 m \omega^2 \rho r^3 {\rm Re}\Big(i \xi_r^* \xi_\perp \Big).
\end{align}

In the wave zone, the fluid displacement consists of an equilibrium (\textquotedblleft non-wave\textquotedblright) component and a dynamical (wave) component, $\bxi = \bxi^{\rm eq} + \bxi^{\rm dyn}$. Since the equilibrium tide component is purely real (assuming negligible dissipation of the equilibrium tide), ${\rm Re} \Big(i \xi_r^{{\rm eq}^*} \xi_\perp^{\rm eq} \Big)=0$, and the equilibrium tide does not contribute to angular momentum transfer. The cross terms ${\rm Re} \Big(i \xi_r^{{\rm dyn}^*} \xi_\perp^{\rm eq} \Big)$, and ${\rm Re} \Big(i\xi_r^{{\rm eq}^*} \xi_\perp^{\rm dyn} \Big)$ are opposed by a nearly equal and opposite Reynold's stress term (see Goldreich \& Nicholson 1989) and do not contribute significantly to angular momentum transfer. Thus, the ${\rm Re} \Big(i \xi_r^{{\rm dyn}^*} \xi_\perp^{\rm dyn} \Big)$ term dominates angular momentum transfer.\footnote{It can be shown that in the WKB limit the Reynold's stress associated with the dynamical response is negligible.} Equation (\ref{Lz2}) then becomes
\begin{align}
\label{Lz5}
\dot{J}_z(r) = & 2 m \omega^2 \rho r^3 {\rm Re}\Big[ i \xi_r^{{\rm dyn}^*} \xi_\perp^{\rm dyn}\Big].
\end{align}

In the outer layers of the WD where $\bxi^{\rm dyn}$ is a pure outgoing wave ($\propto e^{-i k r}$), equation (\ref{xir'}) can be rearranged to obtain the relationship between $\xi_\perp^{\rm dyn}$ and $\xi_r^{\rm dyn}$ in the WKB approximation ($k \gg 1/H$) with $\omega^2 \ll L_l^2$:
\be
\xi_\perp^{\rm{dyn}} \simeq -i \frac{k r}{l(l+1)} \xi_r^{\rm{dyn}}. 
\label{xitscale}
\ee
Then the angular momentum flux is 
\begin{align}
\label{Lscale}
\dot{J}_z & \simeq  2 m l(l+1) \frac{\omega^2 \rho r^2}{k} |\xi_\perp^{\rm dyn}|^2 \nonumber \\
& \simeq \frac{4 \sqrt{l(l+1)} \omega^3 \rho r^3}{N} |\xi_\perp^{\rm dyn}|^2.
\end{align}
where we have used the dispersion relation (equation \ref{k2approx}) with $\omega^2 \ll N^2$ and set $m=2$. This expression agrees with that found in Goldreich \& Nicholson (1989). From the scaling of $\xi_\perp^{\rm dyn }$ provided in equation (\ref{xiperpoutbc}), we see that the angular momentum flux is constant (independent of radius) in the outer layers of the star. Since the wave pattern frequency (in the inertial frame) is $\Omega$ (the orbital frequency), the energy flux carried by the wave is given by $\dot{E} = \Omega \dot{J}_z$. 

Once we have solved our differential equations (\ref{xir'} and \ref{p'}) with the appropriate boundary conditions, we can use equation (\ref{Lz5}) to determine where angular momentum and energy are added to the wave, i.e., where the wave is generated. In the WD interior, the waves travel both inwards and outwards and thus carry no net angular momentum, so the value of $\dot{J}_z$ oscillates around zero. However, near the outer boundary, the value of $\dot{J}_z$ is constant and positive because there only exists an outgoing wave. The region where the value of $\dot{J}_z$ rises to its constant value is the region of wave excitation, because it is in this region where energy and angular momentum are added to the waves (see Section \ref{WDModel}).

The energy and angular momentum carried by the outgoing wave is deposited in the outer envelope of the star. Thus, the constant values of $\dot{J}_z$ and $\dot{E}$ near the outer boundary represent the net angular momentum and energy transfer rates from the orbit to the WD. Since $\xi_\perp^{\rm dyn} \propto M'/a^3$, the angular momentum and energy transfer rates can be written in the form 
\be
\label{edotf}
\dot{J}_z = T_0 F(\omega), \qquad \dot{E} = T_0 \Omega F(\omega), 
\ee
where 
\be
\label{T0}
T_0\equiv G\left({M'\over a^3}\right)^2R^5,
\ee 
and $F$ is a dimensionless function of the tidal frequency $\omega$ and the internal structure of the star. For WDs with rotation rate $\Omega_s$, the tidal frequency is $\omega=2(\Omega-\Omega_s)$.

\section{White Dwarf Models}

\begin{figure*}
\begin{centering}
\includegraphics[scale=.6]{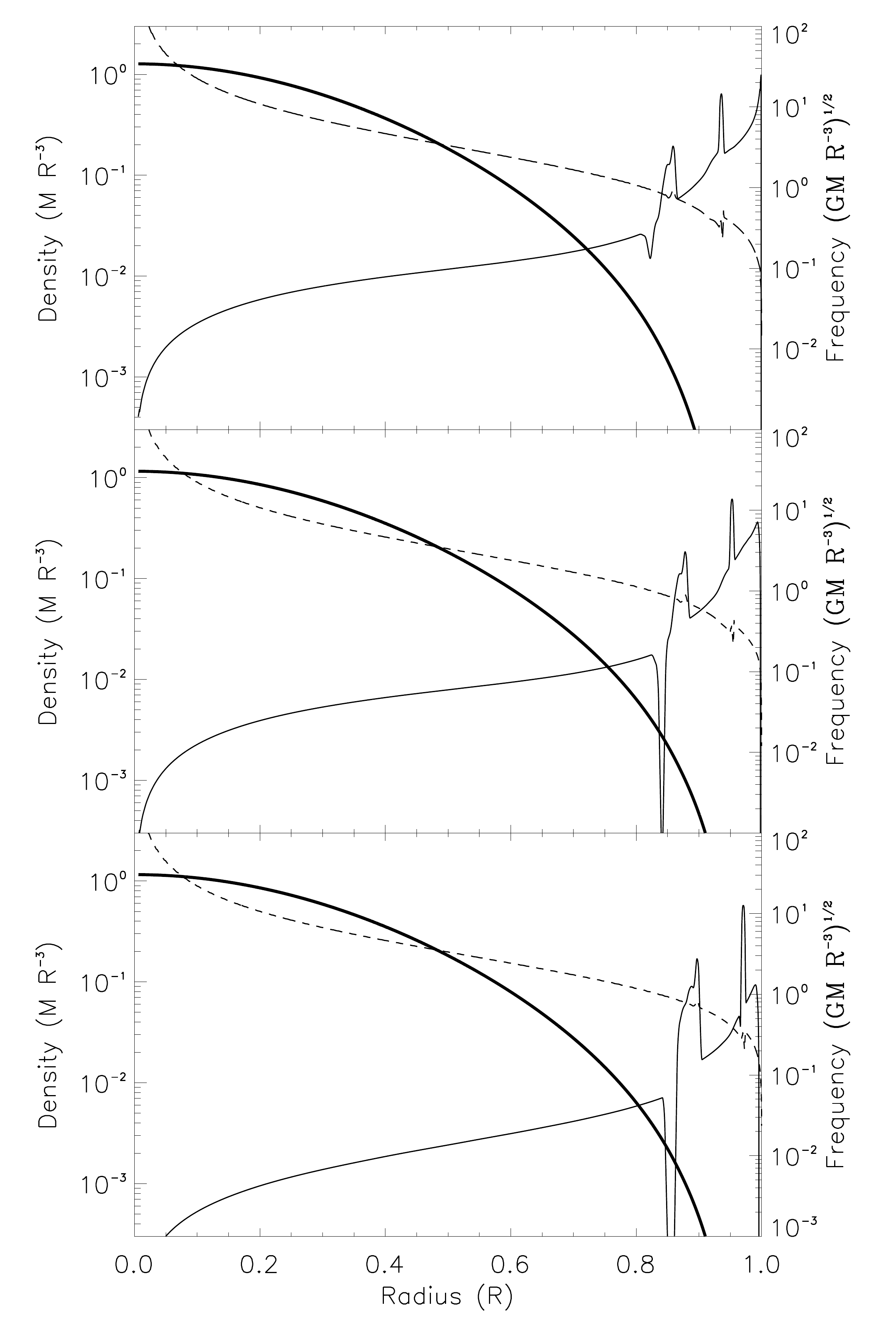}
\caption{\label{WD} The square of the Br\"unt-V\"ais\"al\"a (thin solid lines) and Lamb (dashed lines) frequencies (for $l=2$), in units of $GM/R^3$, and the density (thick solid line) as a function of normalized radius in three WD models. The models are taken from an evolutionary sequence of a DA WD with $M=0.6M_\odot$, $R=8.97 \times 10^3$ km, and effective temperatures of $T=10800$K (top), $T=6000$K (middle), and $T=3300$K (bottom). The spikes in the Br\"unt V\"ais\"al\"a frequency are caused by the composition changes from carbon to helium, and from helium to hydrogen, respectively. Note the formation of a convective zone just below the carbon-helium transition zone as the WD cools.}
\end{centering}
\end{figure*}

Figure \ref{WD} depicts three WD models provided by G. Fontaine (see Brassard et al. 1991). These WD models are taken from an evolutionary sequence of a $M= 0.6 M_\odot$ WD, at effective temperatures of $T=10800$K, $T=6000$K, and $T = 3300$K. The WD has a radius $R \simeq 8.97\times10^8 \textrm{cm}$ and a carbon-oxygen core surrounded by a $10^{-2} M$ helium layer, which in turn is surrounded by a $10^{-4} M$ layer of hydrogen. The models shown have been slightly altered in order to ensure thermodynamic consistency (see Section \ref{error}). 

The Br\"unt-V\"ais\"al\"a frequency can be expressed as
\be
N^2 = \frac{g^2 \rho}{P}\frac{\chi_T}{\chi_P} \big(\nabla_{\rm ad} -\nabla + B\big),
\label{Nbrassard}
\ee
where the symbols have their usual thermodynamic definitions, and the Ledoux term $B$ accounts for composition gradients (see Brassard et al. 1991). In the core of the WD, the value of $N^2$ is very small due to the high degeneracy pressure, which causes $\chi_T$ in equation (\ref{Nbrassard}) to be small. The sharp spikes in $N^2$ are created by the carbon-helium and helium-hydrogen transitions, and are characteristic features of WD models. These sharp features in realistic WDs make it difficult to construct toy WD models or to understand how gravity waves propagate through the WD. From Figure \ref{WD}, it is evident that cooler WDs have smaller values of $N^2$ throughout their interiors. However, the spikes in $N^2$ have little dependence on WD temperature because they are produced by composition gradients rather than thermal gradients, thus these features are unlikely to be strongly affected by tidal heating.

\section{Numerical Calculations of Tidal Response}

\subsection{Numerical Method and Importance of Self-Consistent Stellar Model}
\label{error}

To calculate the amplitude of the gravity waves excited in a WD by its companion, we integrate the inhomogeneous equations (\ref{xir'}) and (\ref{p'}) with the appropriate boundary conditions given by equations (\ref{xiperpoutbc}) and (\ref{xirinbc}). We use the relaxation method discussed by Press et al. (2007). The integration requires a grid of points containing stellar properties ($\rho$, $N^2$, $a_s^2$, $g$) as a function of radius, and solves the equations on a grid of (possibly identical) relaxation points. 

When creating the grid of data points representing the stellar structure, one must be very careful in ensuring that the stellar properties are consistent with one another. In particular, the Br\"unt V\"ais\"al\"a frequency is given by
\be
\label{N}
N^2 = -g\bigg( \frac{\rho'}{\rho} + \frac{g}{a_s^2}\bigg).
\ee
If the value of $N^2$ in our stellar grid is not exactly equal to the right hand side of the above equation as calculated from the values of $\rho$, $g$, and $a_s^2$, the stellar properties will not be self-consistent. Such inconsistency may arise from the inaccuracy of the stellar grids, or from the interpolation of the stellar grids (even if the original grids are exactly self-consistent). We have found that even a small inconsistency can lead to large error in the computed wave amplitude. The reason for this can be understood by examining equation (\ref{yapproxeq}). When tracing back to equations (\ref{xir'}) and (\ref{p'}), one can see that the $N^2$ term on the right-hand side is actually the sum of two terms. That is, the value of $N^2$ on the right-hand side of equation (\ref{zapproxeq}) is calculated via equation (\ref{N}) from our grid of $\rho$, $a_s^2$, and $g$ values, while the $N^2$ term on the left hand side of the equation is taken directly from our grid of $N^2$ values. If these two values of $N^2$ differ (by the amount $\delta N^2$), then a \textquotedblleft false\textquotedblright excitation term will be introduced on the right-hand side of equation (\ref{zapproxeq}), given by
\be
\label{Vf}
V_{f} \simeq - \frac{l(l+1)\delta N^2}{r^2 \omega^2}\frac{U}{g}.
\ee
This false term can vary rapidly with radius depending on the error in the stellar grid. In Section \ref{estimate}, we discuss how sharp changes in the excitation term can be responsible for the excitation of the dynamical component of the tidal response. Thus, the false excitation term introduced by even small numerical inconsistencies can cause large errors in calculations of the dynamical tide.

To test our methods, we calculated the tidal response for a simple massive star model. The results are discussed in Appendix \ref{massive}, and are consistent with previous studies of gravity waves in massive stars (e.g., Zahn 1975,1977 and Goldreich \& Nicholson 1989).

\subsection{Calculation with Toy White Dwarf Model}
\label{toywd}

\begin{figure*}
\begin{centering}
\includegraphics[scale=.6]{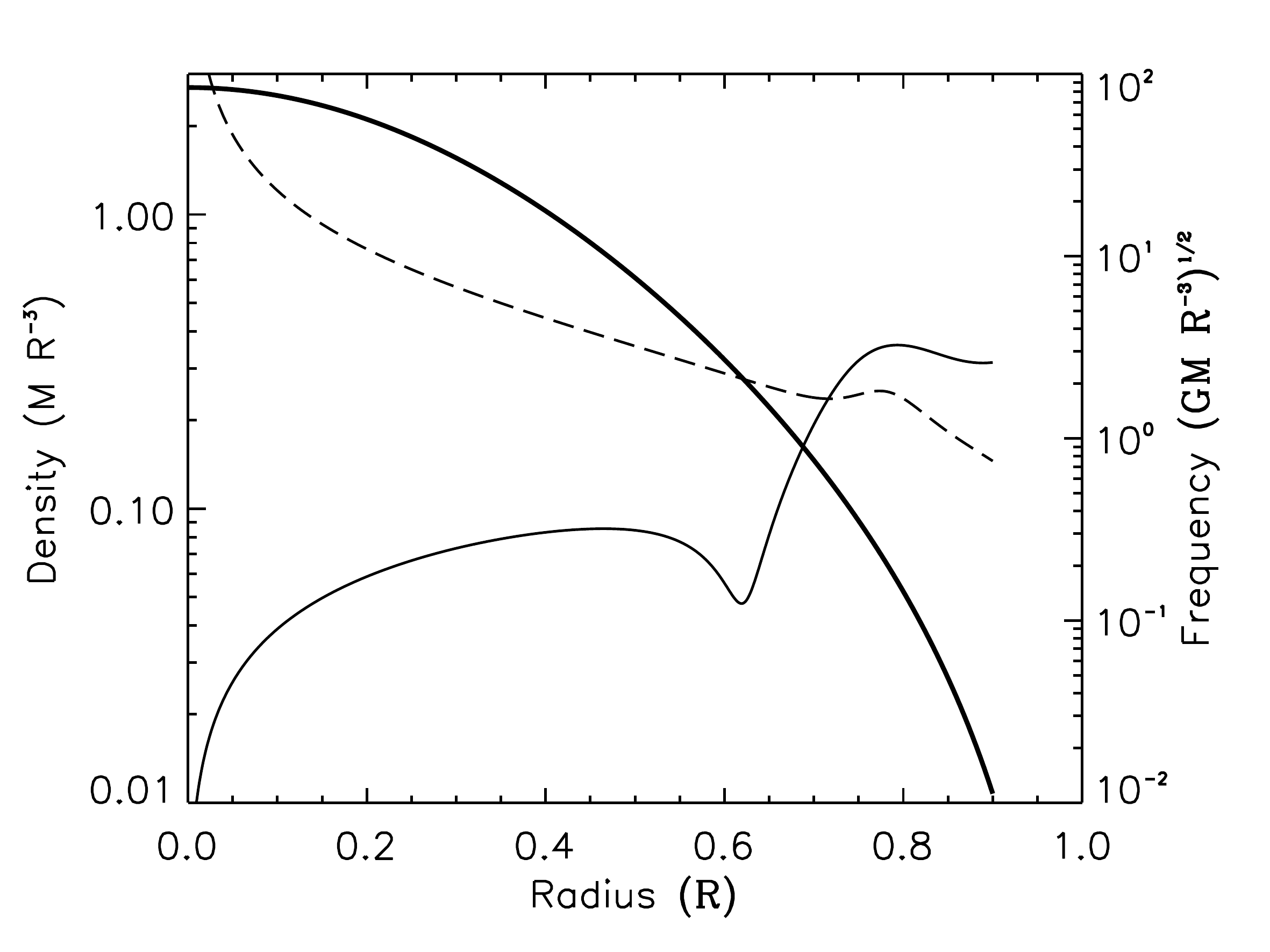}
\caption{\label{poly} The square of the Br\"unt Vais\"al\"a (thin solid line) and Lamb (dashed line) frequencies (for $l=2$), in units of $GM/R^3$, as a function of normalized radius in a toy WD model. Also plotted is the stellar density (thick solid line). The stellar properties are only plotted out to $r= 0.9R$, the location where the outer boundary condition is imposed in the tidal excitation calculation.}
\end{centering}
\end{figure*}

To understand wave excitation in WDs, we first examine a toy model constructed to mimic the structure of a WD.  Examining the $T_{\textrm{eff}} = 10800$K model, we see that it contains a sharp rise in $N^2$ at the carbon-helium boundary, preceded by a small dip in $N^2$ near the top of the carbon layer. Consequently, we have created a toy model with a similar dip and rise in $N^2$ in the outer part of the star. To create this model, we first construct a smooth density profile (identical to that of an $n=2$ polytrope, along with a smooth $N^2$ profile that mimics the dip-rise features associated with the C-He transition in real WDs. Next, we compute a thermodynamically consistent sound speed profile using the equation
\be
\label{N2}
a_s^2 = \bigg(\frac{1}{dP/d\rho} - \frac{N^2}{g^2} \bigg)^{-1}.
\ee
Since the density profile is that of a polytrope, the $dP/d\rho$ term can be calculated analytically. 

\begin{figure*}
\begin{centering}
\includegraphics[scale=.7]{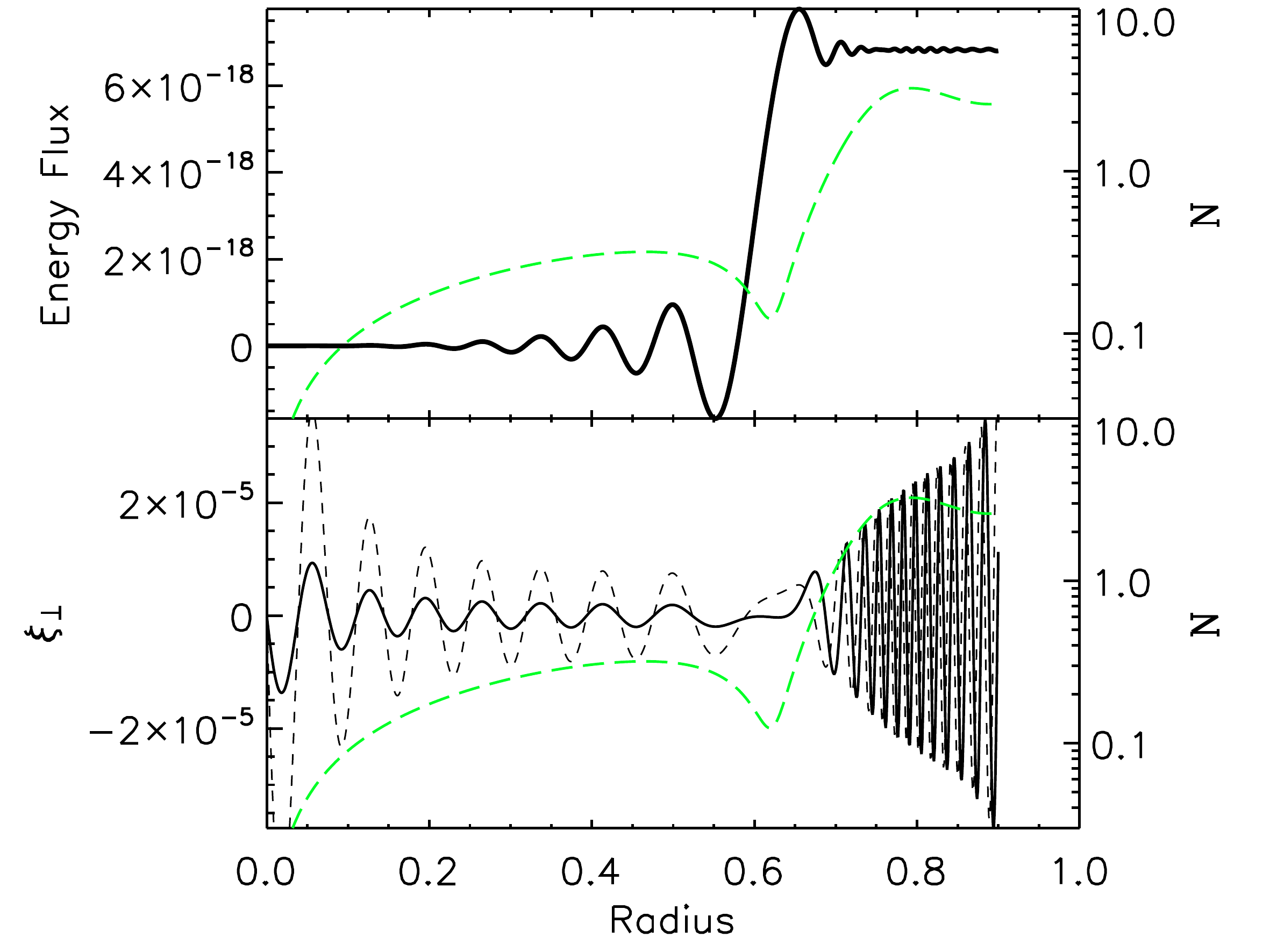}
\caption{\label{PolyRadialEflux} Dynamical tide in a toy WD model (based on the model depicted in Figure \ref{poly}) driven by a companion of mass $M'=M$, with the tidal frequency $\omega=2.3\times10^{-2}$. Top: The energy flux $\dot{E}= \Omega \dot{J}_z$ (dark solid line) as a function of radius, with $\dot{J}_z$ calculated from equation (\ref{Lz5}). All values are plotted in units of $G=M=R=1$. Bottom: The real part of $\xi_\perp^{\rm{dyn}}$ (dark solid line) and imaginary part of $\xi_\perp^{\rm{dyn}}$ (dashed line) as a function of stellar radius. The value of $N^2$ has been plotted (light solid green line) in both panels. In this model, the energy flux rises to its final value near the dip in $N^2$, showing that the wave is excited at this location.}
\end{centering}
\end{figure*}

We solve the forced oscillation equations as a function of the tidal frequency $\omega$. Figure \ref{PolyRadialEflux} shows the energy flux and wave amplitude as a function of radius for a given value of $\omega$. The small oscillations in energy flux are due to imperfect numerical calculation of the dynamical component of the wave and do not actually contribute to energy or angular momentum transfer. We see that waves are excited near the dip of $N^2$ (before $N^2$ rises to a maximum). This is similar to the location of wave excitation in real WD models (see Section \ref{WDModel}). The dip in $N^2$ causes the wave to have a larger wavelength in this region, and so it couples to the companion star's gravitational potential best in this region of the star. Note that although $N^2$ is smaller near the center of the star, no significant wave is excited there since $U \propto r^2$ is negligible. 

\begin{figure*}
\begin{centering}
\includegraphics[scale=.6]{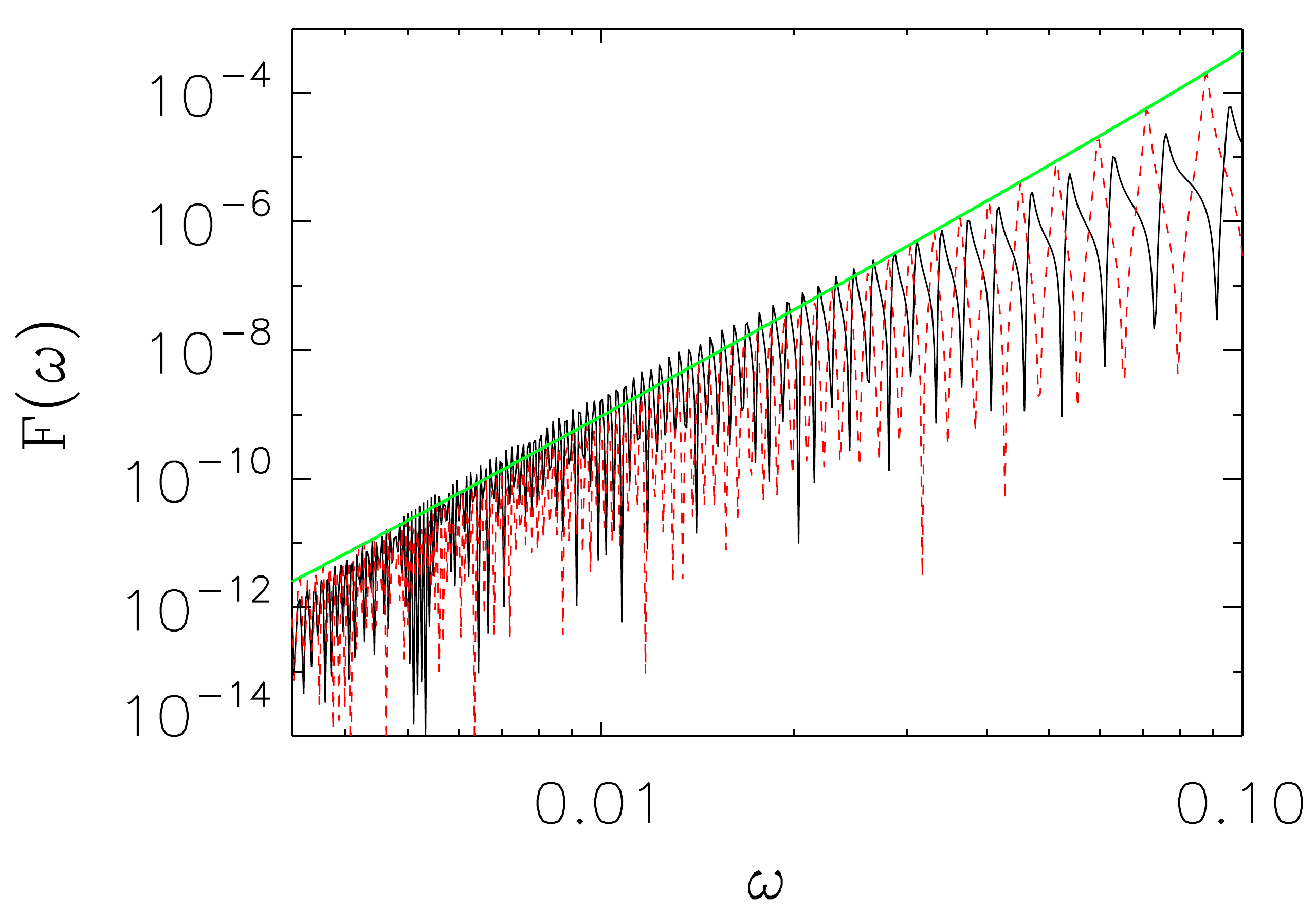}
\caption{\label{EfluxPoly} The dimensionless tidal torque $F(\omega) = \dot{J}_z/T_o$ [see equation (\ref{edotf})] carried by the outgoing gravity wave as a function of the tidal frequency $\omega$ (solid line), for the toy WD model depicted in Figure \ref{poly}. The frequency is in units of $G=M=R=1$. The straight light solid (green) line is calculated from equation (\ref{Efluxapprox2}) and is roughly proportional to $\omega^{5}$. The dashed (red) line is our semi-analytical approximation, with $\alpha=1/5$, $\beta=1/5$, and $\delta=0$ (see Section \ref{estimate}).}
\end{centering}
\end{figure*}

Figure \ref{EfluxPoly} shows a plot of $F(\omega)$. For this model, $F(\omega)$ is not a smooth, monotonic function of $\omega$ as it is for the massive star model studied in Appendix \ref{massive}. Instead, there are many jagged peaks and troughs, causing the value of $F(\omega)$ to vary by two or three orders of magnitude over very small frequency ranges. These features are also present in the real WD models, and will be discussed further in Section \ref{estimate}. Our numerical results indicate that the peaks of $F(\omega)$ can be fitted by $F(\omega) \propto \omega^5$, significantly different from the massive star model.

\subsection{Calculation with Realistic White Dwarf Model}
\label{WDModel}

\begin{figure*}
\begin{centering}
\includegraphics[scale=.7]{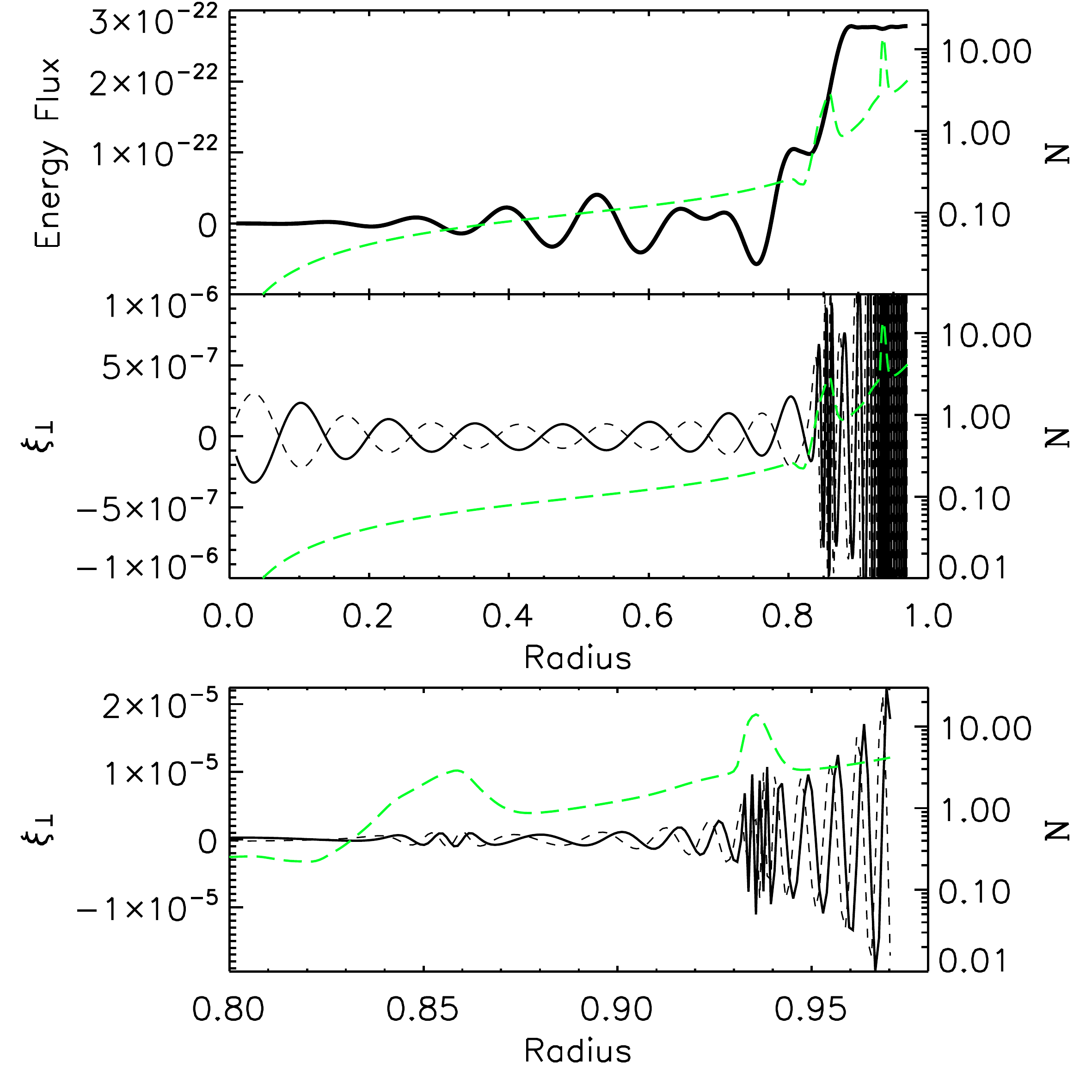}
\caption{\label{WDRadialEflux} Dynamical tide in a realistic WD model (with $M=0.6 M_\odot$, $R= 8.97 \times 10^3$km, and $T_{\rm eff} = 10800$K) driven by a companion of mass $M'=M$, with the tidal frequency $\omega=2\Omega=10^{-2}$. Top: The energy flux $\dot{E} = \Omega \dot{J}_z$ (thick solid line) as a function of radius, calculated from equation (\ref{Lz5}). All values are plotted in units of $G=M=R=1$. Middle: The real part of $\xi_\perp^{\rm{dyn}}$ (solid line) and imaginary part of $\xi_\perp^{\rm{dyn}}$ (dashed line) as a function of stellar radius. Bottom: The same as the middle panel, but zoomed in on the outer layer of the WD. The value of $N^2$ has been plotted as dashed (green) lines in each panel. The energy flux rises to near its final value around the carbon-helium transition region, showing that the wave is excited at this location.}
\end{centering}
\end{figure*}

\begin{figure*}
\begin{centering}
\includegraphics[scale=.7]{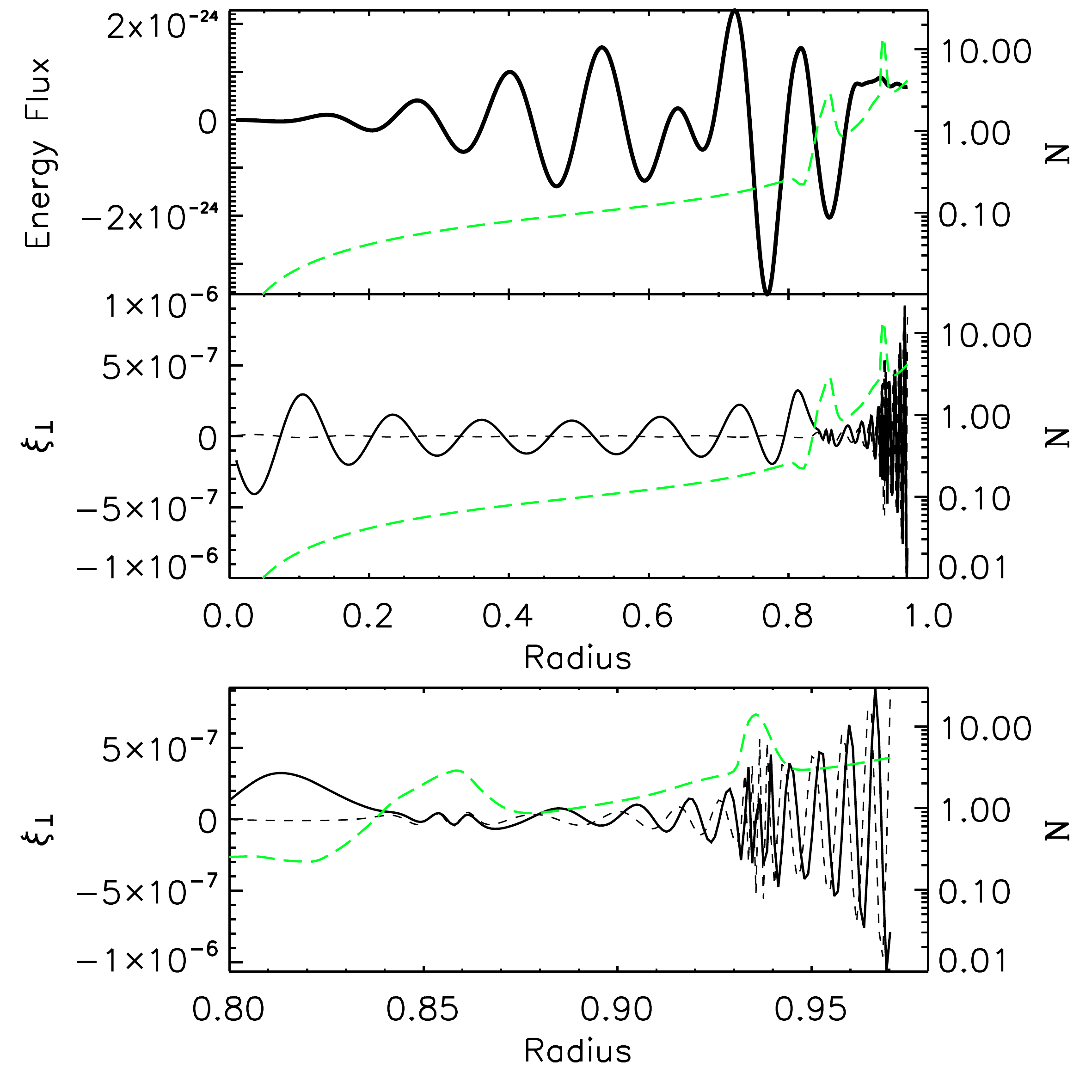}
\caption{\label{WDRadialEflux2} Same as Figure \ref{WDRadialEflux}, but for the tidal frequency $\omega=1.1\times10^{-2}$.}
\end{centering}
\end{figure*}

We now present our numerical results for tidal excitations in realistic WD models. Using the outgoing wave outer boundary condition, we solved the oscillation equations (\ref{xir'}) and (\ref{p'}) for the three WD models described in Section \ref{toywd} (see Figure \ref{WD}). Figures \ref{WDRadialEflux} and \ref{WDRadialEflux2} show plots of the outgoing energy flux as a function of radius for the model with $T_{\rm eff}=10800$K and tidal frequencies of $\omega = 2 \Omega = 10^{-2}$ and $1.1\times10^{-2}$, in units of $G=M=R=1$, respectively. The energy flux jumps to its final value near the carbon-helium transition zone. Once again, the oscillations in energy flux are due to imperfect numerical calculation of the dynamical component of the wave and do not actually contribute to energy or angular momentum transfer. In Figures \ref{WDRadialEflux} and \ref{WDRadialEflux2}, we have smoothed the value of the energy flux to minimize the amplitude of the unphysical oscillations. Note that although Figure \ref{WDRadialEflux2} corresponds to a larger tidal frequency, the outgoing energy flux is about 100 times less than in Figure \ref{WDRadialEflux}. Thus, as in our toy WD model (see Section \ref{toywd}), the tidal energy flux is not a monotonic function of tidal frequency as it is for early-type stars (see Section \ref{massive}).

\begin{figure*}
\begin{centering}
\includegraphics[scale=.6]{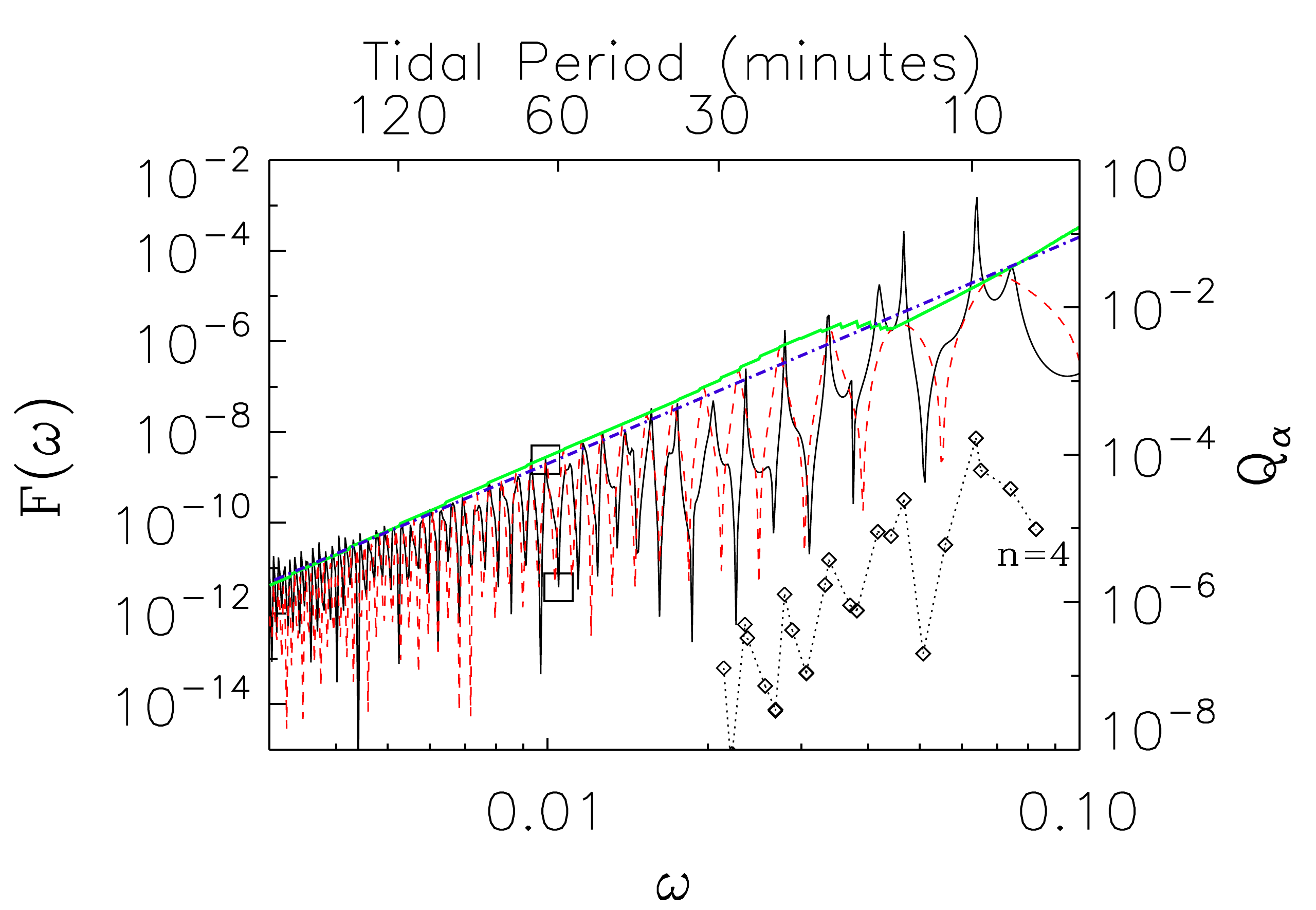}
\caption{\label{WD0090Eflux} The dimensionless tidal torque $F(\omega)= \dot{J}_z/T_o$ [see equation (\ref{edotf})] carried by outgoing gravity waves as a function of the tidal frequency $\omega$ for our WD model with $T_{\rm eff}=10800$K. The two boxed points correspond to $\omega=10^{-2}$ and $1.1\times 10^{-2}$, as depicted in Figures \ref{WDRadialEflux} and \ref{WDRadialEflux2}. The dashed (red) line is our semi-analytical approximation [see equation (\ref{Efluxapprox})], with $\alpha=1/5$, $\beta=1/5$, and $\delta=0$. The smooth solid line corresponds to the maximum values of $F(\omega)$ in our semi-analytical equation, and is calculated from equation (\ref{Fmax}). The dot-dashed (blue) line corresponds to $F(\omega)=20\hat{\omega}^5$ (see Section \ref{spin}). The diamonds connected by the dotted line are the tidal overlap integrals $Q_\alpha$ associated with nearby gravity modes, and the $n=4$ mode is the highest frequency mode shown. The frequency and $Q_\alpha$ are plotted in units of $G=M=R=1$. }
\end{centering}
\end{figure*}

\begin{figure*}
\begin{centering}
\includegraphics[scale=.6]{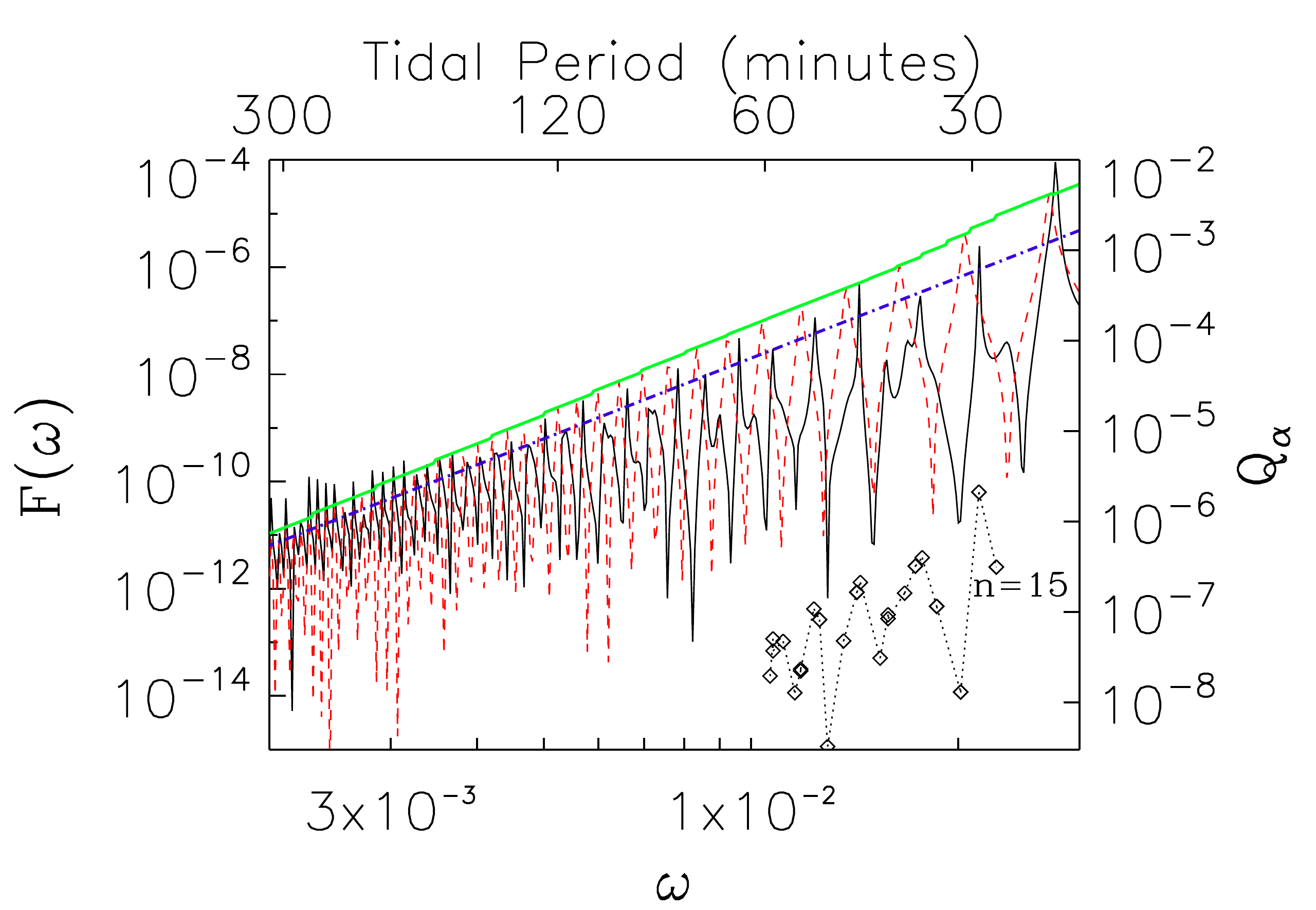}
\caption{\label{WD0130Eflux} Same as Figure \ref{WD0090Eflux}, except for the $T_{\rm eff}=6000$K WD model. In this plot, the dot-dashed (blue) line corresponds to $F(\omega)=200\hat{\omega}^5$. The $n=15$ mode is the highest frequency g-mode shown.}
\end{centering}
\end{figure*}

\begin{figure*}
\begin{centering}
\includegraphics[scale=.6]{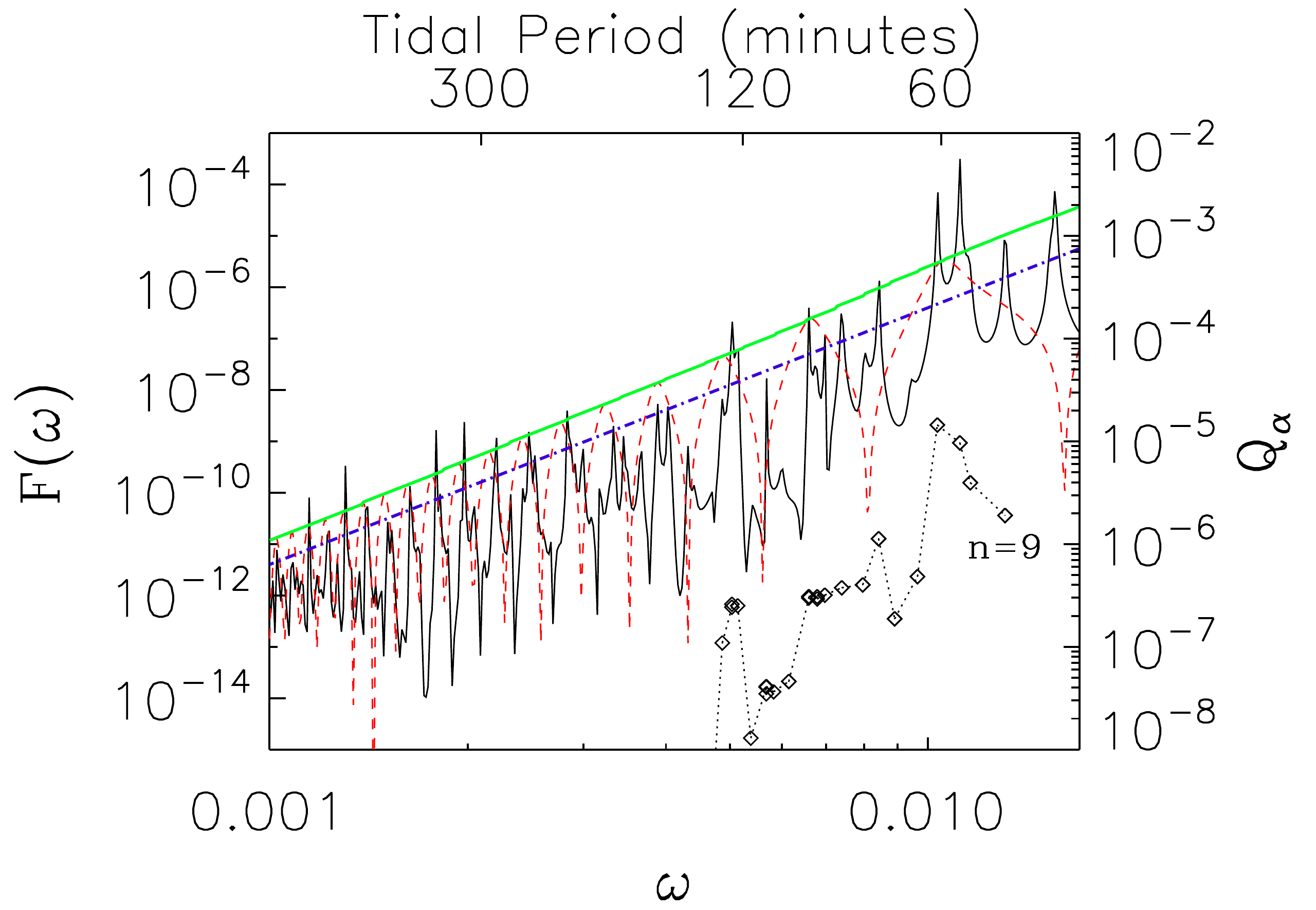}
\caption{\label{WD0160Eflux}Same as Figure \ref{WD0090Eflux}, except for the $T_{\rm eff}=3300$K WD model. In this plot, the dot-dashed (blue) line corresponds to $F(\omega)=4\times10^3\hat{\omega}^5$. The $n=9$ mode is the highest frequency g-mode shown.}
\end{centering}
\end{figure*}

We have calculated the dimensionless tidal torque $F(\omega)= \dot{J}_z/T_o$ [see equation (\ref{edotf})] as a function of $\omega$ for the three WD models depicted in Figure \ref{WD}. The results are shown in Figures \ref{WD0090Eflux}, \ref{WD0130Eflux}, and \ref{WD0160Eflux}. In general, $F(\omega)$ exhibits a strong and complicated dependence on $\omega$, such that a small change in $\omega$ leads to a very large change in $F(\omega)$ (see also Figures \ref{WDRadialEflux}-\ref{WDRadialEflux2}). This dependence is largely due to \textquotedblleft resonances\textquotedblright between the radial wavelength of the gravity waves and the radius of the carbon core, as discussed in Section \ref{estimate}. We also find that the local maxima of $F(\omega)$ can be approximately fitted by the scaling $F(\omega) \propto \omega^5$, similar to the tow WD model discussed in Section \ref{toywd}.

\subsection{Relation to Tidal Overlap Integral}

In an attempt to understand the erratic dependence of the tidal energy transfer rate $\dot E$ on the tidal frequency $\omega$, here we
explore the possible relationship between $\dot E$ and the tidal overlap integral. The energy transfer rate to the star due to tidal interactions can be
written as
\be
\dot E= -2\,{\rm Re}\!\int\!d^3x\,\rho\,{\partial\bxi(\br,t)\over\partial t}\cdot
\nabla U^\star(\br,t).
\ee
With $\bxi(\br,t)=\bxi(\br)e^{-i\omega t}$ and $U(\br,t) = U(r)Y_{22}e^{-i \omega t}$ [see equations (\ref{U})-(\ref{U2})],
we have
\be
\dot E=2\omega {GM'W_{22}\over a^3}\,{\rm Im} \bigg[
\int\!d^3x\,\rho\,\bxi(\br)\cdot\nabla (r^2Y_{22}^\star)\bigg].
\ee
We decompose the tidal response $\bxi(\br,t)$ into the superposition 
of stellar oscillation modes (with each mode labeled by the index $\alpha$):
\be
\bxi(\br,t)=\sum_\alpha a_\alpha(t)\bxi_\alpha(\br),
\ee
where the mode eigenfunction $\bxi_\alpha$ is normalized via
$\int\!d^3x\,\rho\,|\bxi_\alpha|^2=1$. Then the mode amplitude
$a_\alpha(t)$ satisfies the equation
\be
\label{modeamp}
\ddot a_\alpha+\omega_\alpha^2\, a_\alpha +\gamma_\alpha\,\dot a_\alpha
={GM'W_{22}Q_\alpha\over a^3} \ e^{-i\omega t},
\ee
where $\omega_\alpha$ is the mode frequency, $\gamma_\alpha$ is the mode
(amplitude) damping rate, and
$Q_\alpha$ is the {\it tidal overlap integral} with mode $\alpha$:
\be
Q_\alpha=\int\!d^3x\,\rho\,\bxi^\star_\alpha(\br)\cdot\nabla (r^2Y_{22}).
\ee
The steady-state solution of equation (\ref{modeamp}) is
\be
a_\alpha(t)={GM'W_{22}Q_\alpha\over a^3(\omega_\alpha^2-\omega^2
-i\gamma_\alpha\omega)}\,e^{-i\omega t}.
\ee
Thus the tidal energy transfer rate to mode $\alpha$ is
\be
\dot E_\alpha=2\omega\left({GM'W_{22}|Q_\alpha|\over a^3}\right)^2
{\gamma_\alpha\omega\over (\omega_\alpha^2-\omega^2)^2+(\gamma_\alpha
\omega)^2}.
\ee

In paper I, we have computed $\omega_\alpha$ and $Q_\alpha$ for
adiabatic g-modes of several WD models used in this paper. The
eigenfunctions of these modes satisfy the ``reflective'' boundary
condition (i.e., the Lagrangian pressue perturbation $\Delta P$
vanishes) at the WD surface. Our result showed that although the mode
frequency $\omega_\alpha$ decreases as the radial mode number $n$
increases (for a given $l=2$), the overlap integral $|Q_\alpha|$ is a
non-monotonic function of $n$ (or $\omega_\alpha$) due to various
features (associated with carbon-helium and helium-hydrogen transitions) in the $N^2$
profile of the WD models. On the other hand, our calculation of the
tidal response $\bxi(\br,t)$ presented in this paper adopts the
radiative outer boundary condition; this implies significant wave damping at
the outer layer of the star. Because of the difference in the outer
boundary conditions, the mode frequency $\omega_\alpha$ (as computed
using the $\Delta P=0$ boundary condition) does not have special
significance. Nevertheless, we may expect that when $\omega=\omega_\alpha$,
the tidal energy transfer is dominated by a single mode ($\alpha$) and 
$\dot E$ is correlated to $|Q_\alpha|^2$. 

In Figures \ref{WD0090Eflux}, \ref{WD0130Eflux}, and \ref{WD0160Eflux}  we show $|Q_\alpha|$ as a function of $\omega_\alpha$ for a number of
low-order g-modes. It is clear that the peaks and troughs of $F(\omega)$ calculated with an outgoing wave outer boundary condition are associated with the peaks and troughs in the value of $|Q_\alpha|^2$. Thus, the peaks in the value of $F(\omega)$ are {\it not} due to resonances with g-modes, but approximately correspond to the tidal frequencies near the \textquotedblleft intrinsic frequencies\textquotedblright \ of the g-modes with large values of $|Q_\alpha|^2$. Note this correspondence between $|Q_\alpha|^2$ and the local peaks of $F(\omega)$ is not precise (as they are calculated using different boundary conditions), as is clear from the $T_{\rm eff} = 3300$K model (Figure \ref{WD0160Eflux}). Another way to understand the erratic dependence of $F(\omega)$ on $\omega$ lies in the quasi-resonance cavity of the carbon core of the WD (see Section \ref{estimate}).

\subsection{Justification of the Outer Boundary Condition}
\label{nonlinear}

\begin{figure*}
\begin{centering}
\includegraphics[scale=.6]{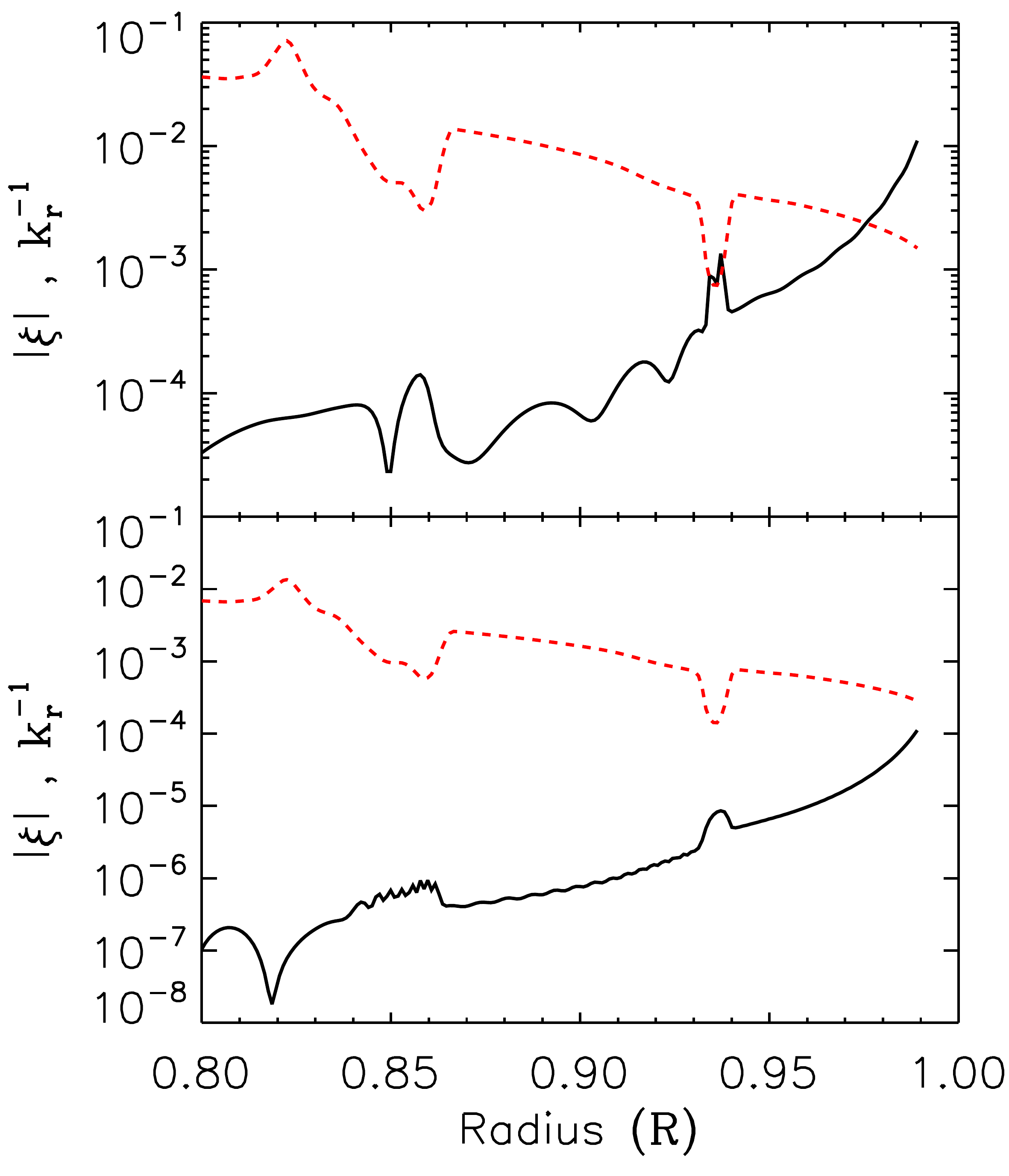}
\caption{\label{nonlin} The magnitude of the gravity wave displacement vector $|\bxi^{\rm dyn}|$ (solid line) as a function of radius for a tidal frequency of $\omega=0.028$ (top) and $\omega=0.0053$ (bottom). Also plotted is the local radial wavelength $k_r^{-1}$ (red dashed line). The wave displacement, wavelength, and frequency are in units where $G=M=R=1$.}
\end{centering}
\end{figure*}

Our calculations in this paper adopt the outgoing wave boundary condition near the stellar surface. This implicitly assumes that gravity waves are absorbed in the outer layer of the WD due to nonlinear effects and/or radiative damping. To analyze the validity of this assumption, we plot the magnitude of the displacement, $|\bxi^{\rm dyn}|$, as a function of radius  in Figure \ref{nonlin}. We have shown the results for tidal frequencies of $\omega=2\Omega=0.028$ and $0.0053$ (corresponding to frequencies near the peaks in $F(\omega)$ shown in Figure \ref{WD0090Eflux}) for our WD model with $T_{\rm eff}=10800$K. We have also plotted the local radial wavelength $k_r^{-1}$ because we expect nonlinear wave breaking to occur when $|\bxi^{\rm dyn}| \gtrsim k_r^{-1}$. 

It is evident from Figure \ref{nonlin} that at relatively high tidal frequencies, the gravity waves become nonlinear in the outer layer of the star, justifying our outgoing wave boundary condition. In some cases, the waves formally reach nonlinear amplitudes ($k_r^{-1}|\xi^{\rm dyn}|>1$) in the helium-hydrogen transition region (demarcated by the dip in $k_r^{-1}$ at $r \simeq 0.935$). This implies that waves may be partially reflected at the helium-hydrogen transition region, although nonlinear damping may also occur before the waves make it to the outermost layers of the WD. The lower frequency gravity waves do not formally reach nonlinear amplitudes in the region depicted in Figure \ref{nonlin}. However, when extending to the lower-density region near the stellar surface, the wave amplitudes will increase further and nonlinearity will set in, although partial reflection may occur due to the shallow convection zone very near the stellar surface. Also note that our calculations are for $\Omega_s=0$. If the WD has a non-negligible spin ($\Omega_s$), a given tidal frequency $\omega=2(\Omega-\Omega_s)$ would correspond to a higher orbital frequency $\Omega$, further increasing the wave amplitudes compared to those shown in Figure \ref{nonlin}. Furthermore, lower frequency waves may damp efficiently via radiative diffusion near the stellar surface. We therefore expect our outgoing wave outer boundary condition to be a good approximation for the frequencies considered in this paper for our warmest WD model.

Our cooler WD models with $T_{\rm eff}=6000$K and $T_{\rm eff}=3300$K do not formally reach the same nonlinear amplitudes as our warmest model. The cooler models have smaller Brunt-Vaisala frequencies, particularly in their outer layers, as can be seen in Figure \ref{WD}. Consequently, the gravity waves have smaller displacements (recall the WKB scaling $\xi_\perp^{\rm dyn} \propto N^{1/2}$ for a constant $\dot{J}_z$) and larger wavelengths (recall $k_r \propto N$). Therefore, gravity waves are less likely to damp due to nonlinear effects in our cooler models, and our outgoing wave outer boundary condition may not be justified at all frequencies considered. More detailed analyses of the nonlinear effects in dynamical tides are necessary (e.g., Barker \& Ogilvie 2010, Weinberg et al. 2011).

\section{Simple Model for Gravity Wave Excitation: Analytical Estimate}
\label{estimate}

\begin{figure*}
\begin{centering}
\includegraphics[scale=.6]{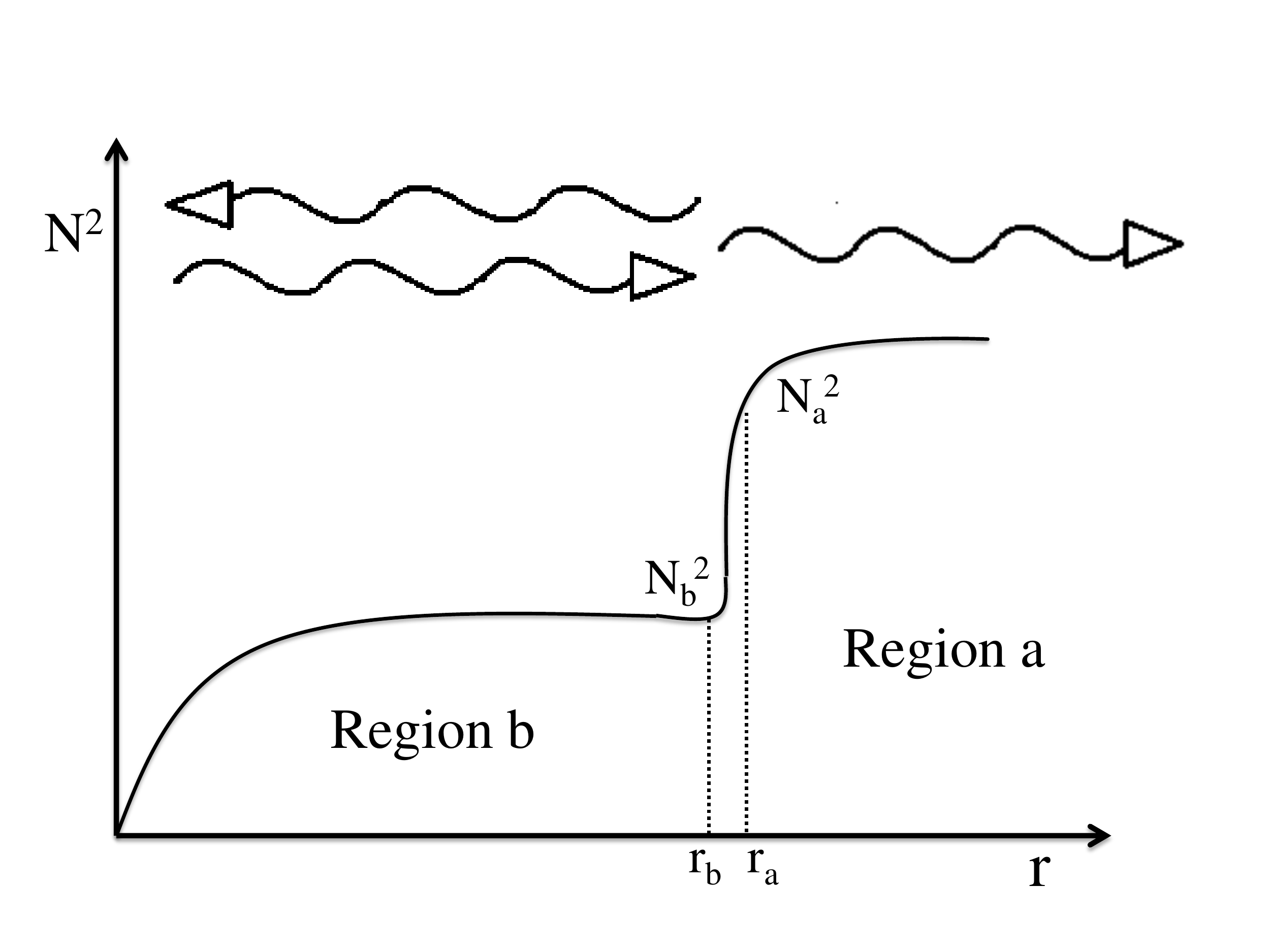}
\caption{\label{region} A diagram showing a simplified model of a white dwarf used in our analytical estimate for gravity wave excitation. The arrows indicate that region b contains both an inward and outward propagating wave, while region a contains only an outward propagating wave.}
\end{centering}
\end{figure*}

To understand our numerical result for the tidal energy transfer rate
$\dot E$ (Section \ref{WDModel}), particularly its dependence on the tidal
frequency $\omega$, here we consider a simple stellar model that, we
believe, captures the essential physics of tidal excitation of gravity
waves in binary WDs. In this model, the star consists of two regions
(see Figure \ref{region}): the outer region with $r>r_a$ (region a)
and the inner region with $r<r_b$ (region b). In each region, the
stellar profiles are smooth, but $N^2$ jumps from $N_b^2$ 
at $r=r_b$ to $N_a^2$ (with $N_a^2 \gg N_b^2)$ at $r=r_a$. The tidal frequency $\omega$
satisfies $\omega^2\ll N_b^2$. As we will see, although waves can propagate in both regions, the sharp jump in $N^2$ makes the inner region behave like a resonance cavity--this is ultimately responsible for the erratic dependence of $F(\omega)$ on the tidal frequency $\omega$.

We start from the wave equation (\ref{z}) for $Z(r)=\chi^{-1/2}r^2\xi_r$:
\be
Z''+k^2(r)Z=V(r),
\ee
with 
\begin{align}
\label{deltak}
k^2(r) &= {l(l+1)N^2\over r^2\omega^2} + \Delta k^2(r) \nonumber \\
 &= {l(l+1)N^2\over r^2\omega^2} \bigg\{1 + \mathcal{O} \bigg[\frac{r^2}{H^2}\frac{\omega^2}{l(l+1)N^2}\bigg]\bigg\}
\end{align}
\be
\label{deltaV}
V(r)=-\chi^{-1/2}{l(l+1)N^2\over \omega^2}{U\over g}
\left[1-{2r\over H}{\omega^2\over l(l+1)N^2}\right],
\ee
where $H=a_s^2/g(\lo r)$ is the pressure scale height.
The above expressions are valid in both regions of the star,
and we have assumed $\omega^2\ll L_l^2$ and $\omega^2\ll N^2$
[more general expressions are given by equations (\ref{k}) and (\ref{V})].
Since the stellar profiles are smooth in each of the two regions,
the non-wave (``equilibrium'') solution is given by
\be
Z^{\rm eq}(r)\simeq {V\over k^2}-{1\over k^2}\left({V\over k^2}\right)''
=Z_0+\Delta\!Z,
\ee
where 
\begin{align}
\label{Z0}
& Z_0=-\chi^{-1/2}{r^2U\over g},\\
& \Delta\!Z=Z_0{\beta \over k^2H^2},
\end{align}
and $\beta$ is a constant (with $\beta\sim 1$).\footnote{From equation (\ref{k}), we find that $\Delta k^2$ in equation (\ref{deltak}) is given by $\Delta k^2 = (H_\rho^{-1})'/2 - (2H_\rho)^{-2} + H_p^{-1}[-(\ln H_p)' + H_\rho^{-1} - H_p^{-1}]$, where $H_p = H$ and $H_\rho=-\rho'/\rho$. In the isothermal region, $H_\rho=H$, and we have $\Delta k^2 = -(2 H_\rho)^{-2}$. In the region satisfying $P \propto \rho^{5/3}$, we have $H_\rho = (5/3) H$, and $\Delta k^2 \simeq 1/(12 H_\rho^2)$. Thus the parameter $\beta$ in equation (\ref{Z0}) ranges from $|\beta| \lesssim 0.1$ to $0.3$.} Note that the above
solution for $Z^{\rm eq}$ breaks down around $r=r_{\rm in}$ 
(where $\omega^2=N^2$). At distances sufficiently far away from $r_{\rm in}$,
we have $k\gg 1/H$.

The general solution to equation (\ref{z}) consists of the non-wave part
$Z^{\rm eq}$ and the wave part $Z^{\rm dyn}$. In region b 
there exist both ingoing and outgoing waves. Thus
\begin{align}
Z(r)=& Z^{\rm eq}(r)+A_+ \exp \bigg(i\!\int_{r_{\rm in}}^r\!\!k\,dr\bigg) \nonumber \\
&+A_- \exp \bigg(-i\!\int_{r_{\rm in}}^r\!\!k\,dr \bigg)
\end{align}
for $r_{\rm in}<r<r_b$,
where $A_+$ and $A_-$ are slow-varying functions of $r$.
In region a, we require there be no ingoing wave. Thus for
$r>r_a$, 
\be
Z(r)=Z^{\rm eq}(r)+A \exp \bigg(-i\!\int_{r_0}^r\!\!k\,dr \bigg)
\ee
where $r_0>r_a$ is a constant, and $A$ varies slowly with $r$.
Note that $Z^{\rm eq}$ is discontinuous between the two regions.

At the inner boundary $r=r_{\rm in}$, gravity waves are perfectly reflected.
Thus we have $A_-=-e^{i\delta}A_+$, where $\delta$ is a constant 
phase that depends on the details of the disturbance around and inside $r_{\rm in}$.
To determine $A$ and $A_+$ we must match the solutions in the two 
regions. Although in reality $r_a$ is somewhat larger than $r_b$,
we shall make the approximation $r_a\simeq r_b$, and label the physical
quantities on each side with the subscript ``a'' or ``b''. Note that
$Z^{\rm eq}_a-Z^{\rm eq}_b \simeq -\Delta Z_b$ since $k_a^2\gg k_b^2$,
and $(dZ^{\rm eq}/dr)_a-(dZ^{\rm eq}/dr)_b\simeq -(\alpha/H)\Delta Z_b$
where $\alpha$ is a constant ($|\alpha|\sim 1$). Matching 
$Z$ and $dZ/dr$ across $r=r_b\simeq r_a$, we obtain the expression for the wave amplitude
at $r=r_a$:
\be
 A \exp \bigg(\!-i\!\int_{r_{0}}^{r_a}\!\!k\,dr\! \bigg)
=\Delta Z_b\left[ {1-(\alpha/k_bH)\tan\varphi\over 1+i(k_a/k_b)\tan\varphi}
\right],
\ee
which entails
\be
\label{amag}
|A|=|\Delta Z_b|\, 
{\left| 1-(\alpha/k_bH)\tan\varphi\right|
\over \left[1+(k_a/k_b)^2\tan^2\varphi\right]^{1/2}},
\ee
where 
\be
\varphi=\int_{r_{0}}^{r_b}\!\! k\,dr-{\delta\over 2}.
\ee
Clearly, $|A|$ reaches the maximum $|\Delta Z_b|$ at $\varphi=0$,
and $|A|\simeq |\Delta Z_b (\alpha/k_a H)|$ at $\varphi=\pi/2$.

The Lagrangian displacement for the outgoing gravity wave in region a 
is given by
\begin{align}
\xi_\perp^{\rm dyn} & \simeq -{ikr\over l(l+1)}\xi_r^{\rm dyn} \nonumber \\
&=-{ik\chi^{1/2}\over l(l+1)r}A\exp\!\left(\!-i\!\int^r_{r_{0}}\!\! k\,dr\!\right).
\end{align}
The tidal energy transfer rate $\dot E$ 
is equal to the energy flux carried by the wave. Using equation (\ref{Lscale}), we have
\be
\label{Efluxapprox}
\dot E=\Omega \dot J_z=4\Omega k_a |A|^2,
\ee
where $k_a=\sqrt{l(l+1)}N_a/(r_a\omega)$ and $|A|$ are evaluated
at $r=r_a$. Using $|A|_{\rm max}=|\Delta Z_b|$, we obtain the maximum 
tidal energy transfer rate as a function of the tidal frequency $\omega$ 
and the orbital frequency $\Omega$:
\be
\label{Efluxapprox2}
\dot E_{\rm max} \simeq {6\pi\beta^2\over 5}{\rho_a r_a^7 N_a\over
N_b^4\left[l(l+1)\right]^{5/2}\!g_a^2}\!\!\left({r_a\over H_a}\right)^{\!4}\!\!
\left({M'\over M_t}\right)^{\!2}\!\Omega^5\!\omega^5.
\ee
The corresponding dimensionless tidal torque [see equation (\ref{edotf})] is 
\be
\label{Fmax}
F_{\rm max} (\omega) = {6\pi\beta^2\over 5}{G \rho_a r_a^7 N_a\over
N_b^4\left[l(l+1)\right]^{5/2}g_a^2R^5}\left({r_a\over H_a}\right)^4\! \omega^5.
\ee
This scaling [$F(\omega) \propto \omega^5$] agrees with our numerical results for the toy WD models (Section \ref{toywd}) and realistic WD models (Section \ref{WDModel}). 

Realistic WD models are obviously more complicated than the analytical model considered in this section (see Figure \ref{region}). To evaluate the tidal energy transfer rate $\dot{E}$ using equation (\ref{Efluxapprox}) [with $|A|$ given by equation (\ref{amag})] and $\dot{E}_{\rm max}$ using equation (\ref{Efluxapprox2}) for our WD models, we choose $r_b$ at the location where $d\ln N^2/dr$ is largest in the helium-carbon transition region. We then set the location of $r_a$ to be one half of a wavelength above $r_b$, i.e., by finding the location $r_a$ such that the equation $\pi = \int^{r_a}_{r_b} k dr$ is satisfied, where $k$ is given by equation (\ref{k2approx}). For the three models considered in Section 6, we find that $r_a$ thus calculated typically lies near the peak in $N^2$ associated with the carbon-helium transition region. 

In Figures \ref{EfluxPoly}, \ref{WD0090Eflux}, \ref{WD0130Eflux}, and \ref{WD0160Eflux}, we compare the analytical results based on equations (\ref{Efluxapprox}) and (\ref{Efluxapprox2}) to our numerical calculations. We see that the erratic dependence of $F(\omega)$ on the tidal frequency $\omega$ can be qualitatively reproduced by our analytical expression (\ref{Efluxapprox}), and the maximum $F_{\rm max}$ is also well approximated by equation (\ref{Fmax}). Our analytical estimate works best for the WD model with $T_{\rm eff}=10800$K, but it does a poor job of approximating the value of $F(\omega)$ for the WD model with $T_{\rm eff}=3300$K. We attribute this disagreement to the lower value of $N^2$ in the cool WD model because our assumption that $N^2 \gg \omega^2$ is not satisfied. Instead, we find that gravity waves are excited near the spike in $N^2$ associated with the helium-hydrogen transition region in the cool WD model. 

For each model shown in Figures \ref{WD0090Eflux}-\ref{WD0160Eflux}, our model also breaks down at the highest and lowest frequencies shown. These discrepancies are likely related to errors in our numerical methods. At the highest frequencies shown. the approximation $k_r \gg 1/H$ begins to break down, causing error in our outer boundary condition. At the lowest frequencies shown, extremely fine grid resolution is needed to resolve the dynamical component of the tidal response, and so slight thermodynamic inconsistencies may introduce significant errors (see Section \ref{error}).

\section{Spin-Orbit Evolution}
\label{spin}

The tidally-excited gravity waves and their dissipations cause energy
and angular momentum transfer from the orbit to the star, leading to
spin-up of the WD over time.  In this section, we study the spin-orbit
evolution of WD binaries under the combined effects of tidal
dissipation and gravitational radiation.  In general, the tidal torque
on the primary star $M$ from the companion $M'$ and the tidal energy
transfer rate can be written as [see equations (\ref{edotf}) and (\ref{T0})]
\be 
T_{\rm tide}=T_0 F(\omega),\quad 
\dot E_{\rm tide}=T_0\Omega F(\omega),
\label{eq:ttide}\ee
with $T_0 = G(M'/a^3)^2R^5$. In previous sections, we have computed
$F(\omega)$ for various non-rotating ($\Omega_s=0$) WD models
(and other stellar models). To study the spin-orbit evolution, here we
assume that for spinning WDs, the function $F(\omega)$ is the
same as in the non-rotating case. This is an approximation because a
finite $\Omega_s$ can modify gravity waves in the star through
the Coriolis force (gravity waves become the so-called Hough waves) and
introduce inertial waves, which may play a role in the dynamical
tides. In other words, the function $F$ generally depends on not only
$\omega$ but also $\Omega_s$.  However, we expect that when the
tidal frequency $\omega=2(\Omega-\Omega_s)$ is larger than $\Omega_s$,
i.e., when $\Omega\go 3\Omega_s/2$, the effect of rotation on the
gravity waves is small or modest. Also, we assume that the WD exhibits
solid-body rotation, which would occur if different layers of the WD
are strongly coupled (e.g., due to viscous or magnetic stresses).\footnote{In a medium containing a magnetic field, we expect differential rotation to be smoothed out by magnetic stresses on time scales comparable to the Alfven wave crossing time. The Alfven wave crossing time is $t_A = R\sqrt{4\pi\rho}/B \simeq 1 \rm{yr}$ for a magnetic field strength of $B=10^5$gauss and a density of $\rho=10^6$g/cm$^3$. Since the Alfven wave crossing time is always much smaller than the inspiral time for WDs, we expect solid body rotation to be a good approximation.}

Before proceeding, we note that in the weak friction theory of
equilibrium tides (e.g., Darwin 1879; Goldreich \& Soter 1966;
Alexander 1973; Hut 1981), the tidal torque is related to the
tidal lag angle $\delta_{\rm lag}$ or the tidal lag time 
$\Delta t_{\rm lag}$ by
\be
T_{\rm tide}=3k_2T_0\delta_{\rm lag},\quad
{\rm with}~~\delta_{\rm lag}=(\Omega-\Omega_s) \Delta t_{\rm lag},
\ee
where $k_2$ is the Love number. Often, a dimensionless tidal quality factor $Q_{\rm tide}$ is introduced (e.g. Goldreich \& Soter 1966) such that 
$\Delta t_{\rm lag}=1/(|\omega| Q_{\rm tide})$ (valid only for $\omega \neq 0$). Thus, if we use the weak-friction theory to parametrize our dynamical
tide, $F(\omega)$ would correspond to 
\begin{align}
F(\omega) &= 3k_2\delta_{\rm lag} \nonumber \\
&=3k_2(\Omega-\Omega_s)\Delta t_{\rm lag} \nonumber \\
&={3k_2\over 2Q_{\rm tide}}{\rm sgn}(\Omega-\Omega_s).
\end{align}
Obviously, the effective 
$Q_{\rm tide}$ would depend strongly on $\omega$ as opposed to
being a constant (assuming constant lag angle) or being proportional to $1/|\omega|$ (assuming constant lag time, appropriate for a viscous fluid).

With equation (\ref{eq:ttide}) and the assumption in $F(\omega)$, 
the WD spin evolves according to the equation
\be
\dot \Omega_s={T_0 F(\omega)\over  I},
\label{Omegasdot}
\ee
where $I$ is the moment of inertia of the WD ($I\simeq 0.169MR^2$
for our $M=0.6M_\odot$ WD models). The orbital energy 
$E_{\rm orb}=-GMM'/(2a)$ satisfies the equation
\be
\label{Eorbit}
\dot{E}_\textrm{orb} = -\dot{E}_{\textrm{tide}} - \dot{E}_{\textrm{GW}},
\ee
where $\dot{E}_{\textrm{GW}}(>0)$ is the energy loss rate due to 
gravitational radiation. The evolution equation for the orbital angular 
frequency $\Omega=(GM_t/a^3)^{1/2}$ is then 
\be
\dot\Omega={3T_0F(\omega)\over\mu a^2}+{3\Omega\over 2t_{\rm GW}},
\label{Omegadot}
\ee
where $\mu=MM'/M_t$ is the reduced mass of the binary, and 
$t_{\rm GW}$ is the orbital decay time scale ($|a/\dot a|$) due to 
gravitational radiation:
\begin{align}
t_{\rm GW} &= \frac{5c^5}{64G^3}\frac{a^4}{MM'M_t}\nonumber\\
&= 3.2\times 10^{10} \bigg(\frac{M_{\odot}^2}{MM'}\bigg)\bigg(\frac{M_t}
{2M_{\odot}}\bigg)^{\!\!1/3}\!\! \bigg(\!\frac{\Omega}{0.1\,\textrm{s}^{-1}}
\bigg)^{\!\! -8/3}\,{\rm s},
\label{tgw}
\end{align}

%%%%%%%%%%%%
\subsection{Synchronization}

Using our results for the function $F(\omega)$ obtained in previous
sections, we integrate equations (\ref{Omegasdot}) and
(\ref{Omegadot}) numerically to obtain the evolution of the WD
spin. Since at large $a$ (small $\Omega$) the orbital decay time $\sim
t_{\rm GW}\propto \Omega^{-8/3}$ is is much shorter than the time
scale for spin evolution, $t_{\rm spin}=\Omega_s/\dot\Omega_s\propto
1/(\Omega^4 F)$, we start our integration with $\Omega_s\ll \Omega$ at
a small orbital frequency (an orbital period of several hours).

\begin{figure*}
\begin{centering}
\includegraphics[scale=.55]{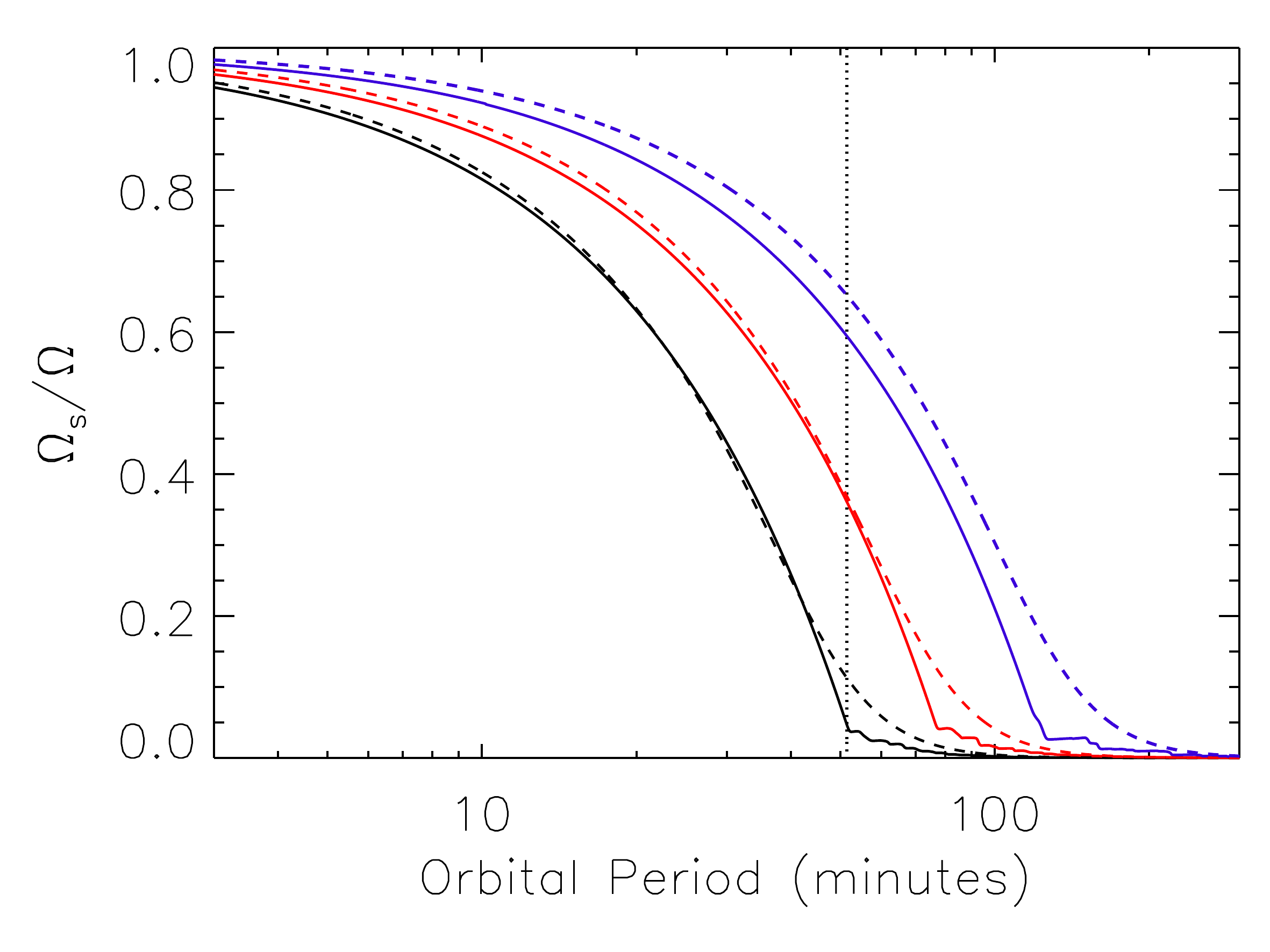}
\caption{\label{spin1} Evolution of the spin frequency $\Omega_s$ in units of the orbital frequency $\Omega$ as a function of the orbital period. The solid black, red, and blue lines correspond to our WD models with $T_{\rm eff}=10800$K, $T=6000$K, and $T=3300$K, respectively. The black, red, and blue dashed lines correspond to evolutions using $F=20\hat{\omega}^5$, $F=200\hat{\omega}^5$, $F=4\times10^3\hat{\omega}^5$, respectively (these functions $F(\omega)$ approximate the like-colored WD models, see Figures \ref{WD0090Eflux}-\ref{WD0160Eflux}). The vertical dotted line denotes the critical orbital period, $2\pi/\Omega_c$ [see equation (\ref{omc})], corresponding to the black dashed line. In these evolutions, $M'=M$ and the WDs initially have $\Omega_s=0$.}
\end{centering}
\end{figure*}

\begin{figure*}
\begin{centering}
\includegraphics[scale=.55]{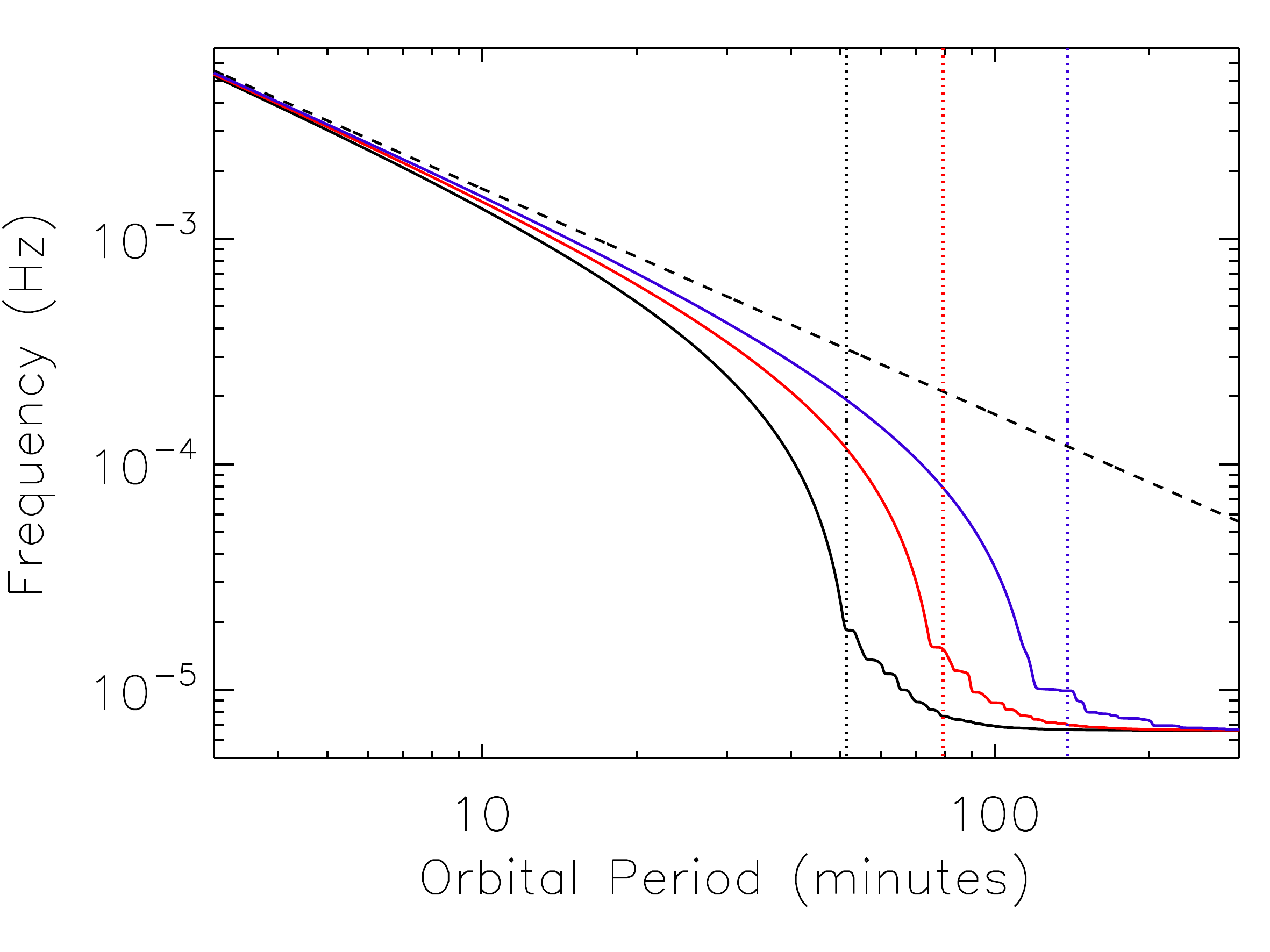}
\caption{\label{spin2} The spin frequency $\Omega_s/(2\pi)$ in units of ${\rm Hz}$ as a function of orbital period. The solid black, red, and blue lines correspond to our WD models with $T_{\rm eff}=10800$K, $T=6000$K, and $T=3300$K, respectively. The dashed line shows the orbital frequency, $\Omega/(2\pi)$. The dotted vertical black, red, and blue lines are the values of $2\pi/\Omega_c$ for $F=20\hat{\omega}^5$, $F=200\hat{\omega}^5$, $F=4\times10^3\hat{\omega}^5$, respectively. In these evolutions, $M'=M$ and the WDs initially have $\Omega_s=\Omega/4$.}
\end{centering}
\end{figure*}

The results for our three WD models are shown in Figures \ref{spin1} and \ref{spin2}.
Note we only include the effects of tides in the primary star ($M$), and treat the
companion ($M'$) as a point mass. All three models have the same
WD masses ($M=M'=0.6M_\odot$), but different temperatures.
Also note that the minimum binary separation
(before mass transfer or tidal disruption occurs) is 
$a_{\rm min}\simeq 2.5(M_t/M)^{1/3}R$, corresponding to
the minimum orbital period
\be
P_{\rm min}\simeq (1.1\,{\rm min})\,M_1^{-1/2}R_4^{3/2},
\ee
where $M_1\equiv M/(1\,M_\odot)$ and $R_4=R/(10^4\,{\rm km})$.
We see that for all models, appreciable spin-orbit synchronization is achieved
before mass transfer or tidal disruption. However, depending 
in the WD temperature, the rates of spin-orbit synchronization 
are different.

The basic feature of the synchronization process can be
obtained using an approximate expression for the dimensionless
function $F(\omega)$. We fit the local maxima of our numerical
results depicted in Figures \ref{WD0090Eflux}-\ref{WD0160Eflux} by the function
\be 
F(\omega)=f\omega^5={\hat f}{\hat\omega}^5,
\ee
where $\hat\omega=\omega/(GM/R^3)^{1/2}$, and ${\hat f}\simeq 20,\,
200,\,4\times 10^3$ for the $T_{\rm eff}=$10800K, 6000K, and 3300K
models, respectively. Suppose $\Omega_s\ll \Omega$ at large
orbital separation. We can define the {\it critical orbital frequency},
$\Omega_c$, at which spinup or synchronization becomes efficient,
by equating $\dot\Omega$ and $\dot\Omega_s$ (with $\Omega_s\ll\Omega$). 
Note that since the orbital decay rate due to tidal energy transfer
[the first term in equation (\ref{Omegadot})] is much smaller than the 
spinup rate $\dot\Omega_s$,
the orbital decay is always dominated by the gravitational radiation,
i.e., $\dot\Omega \simeq 3\Omega/(2t_{\rm GW})$. With 
$T_o = \bar{T}_o \Omega^4$ and $t_{\rm GW} =\bar{t}_{\rm GW} \Omega^{-8/3}$, 
we find
\begin{align}
\Omega_c &\simeq \bigg(\frac{3 I}{64 f\bar{T}_o \bar{t}_{\rm GW}}
\bigg)^{3/16}\nonumber\\
&=\left[{3\kappa\over 5{\hat f}}{M_t^{5/3}\over M'M^{2/3}}
\left({GM\over Rc^2}\right)^{5/2}\right]^{3/16}\left({GM\over R^3}
\right)^{1/2}\nonumber\\
&=(3.8\times 10^{-3}{\rm s}^{-1})\,
\left({\kappa_{0.17}M_{t1}^{5/3}M_1^{9/2}
\over {\hat f}M_1' R_4^{21/2}}\right)^{3/16},
\label{omc}
\end{align}
where $\kappa=0.17\kappa_{0.17}=I/(MR^2)$, $M_1'=M'/(1\,M_\odot)$,
and $M_{t1}=M_t/(1\,M_\odot)$. For $\Omega\lo \Omega_c$, tidal synchronization is inefficient.
For $\Omega \go \Omega_c$, the spinup rate $\dot\Omega_s$ becomes larger
than $\dot\Omega$ and the system will become increasingly synchronized.
In fact, when $\Omega\go\Omega_c$, an approximate analytic
expression for the spin evolution can be obtained
by assuming {\it a posteriori} $(\dot \Omega_s-\dot\Omega)\ll \dot\Omega$.
With $\dot\Omega\simeq 3\Omega/(2t_{\rm GW})\simeq \dot\Omega_s$, 
we find
\begin{align}
\Omega_s &\simeq \Omega - \Omega_c^{16/15} \Omega^{-1/15} \qquad
\big({\rm for} \ \Omega \go \Omega_c \big).
\label{omsapprox}
\end{align}
This expression provides an accurate representation of the numerical
solutions.

Note that we can derive a similar equation as (\ref{omsapprox})
for more general tidal torques. For example, assume
\be
\dot\Omega_s=A\Omega^4(\Omega-\Omega_s)^n,
\ee
where $n$ and $A$ are constants.
With $\dot\Omega=B\Omega^{11/3}$ (where $B$ is a constant) and assuming
$\dot\Omega_s\simeq \dot\Omega$, we find 
\be
\Omega_s\simeq \Omega-\Omega_c \left({\Omega_c\over\Omega}\right)^{1/(3n)},
\label{eq:omegas}\ee
for $\Omega\go\Omega_c$, where 
\be
\Omega_c=\left({B\over A}\right)^{3/(3n+1)}. 
\label{omegac2}
\ee
Note that our equation (\ref{omsapprox})
corresponds to $n=5$, which implies $\Omega-\Omega_s\simeq
\Omega_c$ for $\Omega\go\Omega_c$. By contrast, in the equilibrium tide model (with constant lag time),
$n=1$, so $(\Omega-\Omega_s)$ changes moderately as the orbit decays.

%%%%%%%%%%%%%%%%
\subsection{Tidal Effect on the Orbital Decay Rate and Phase of Gravitational Waves}

\begin{figure*}
\begin{centering}
\includegraphics[scale=.55]{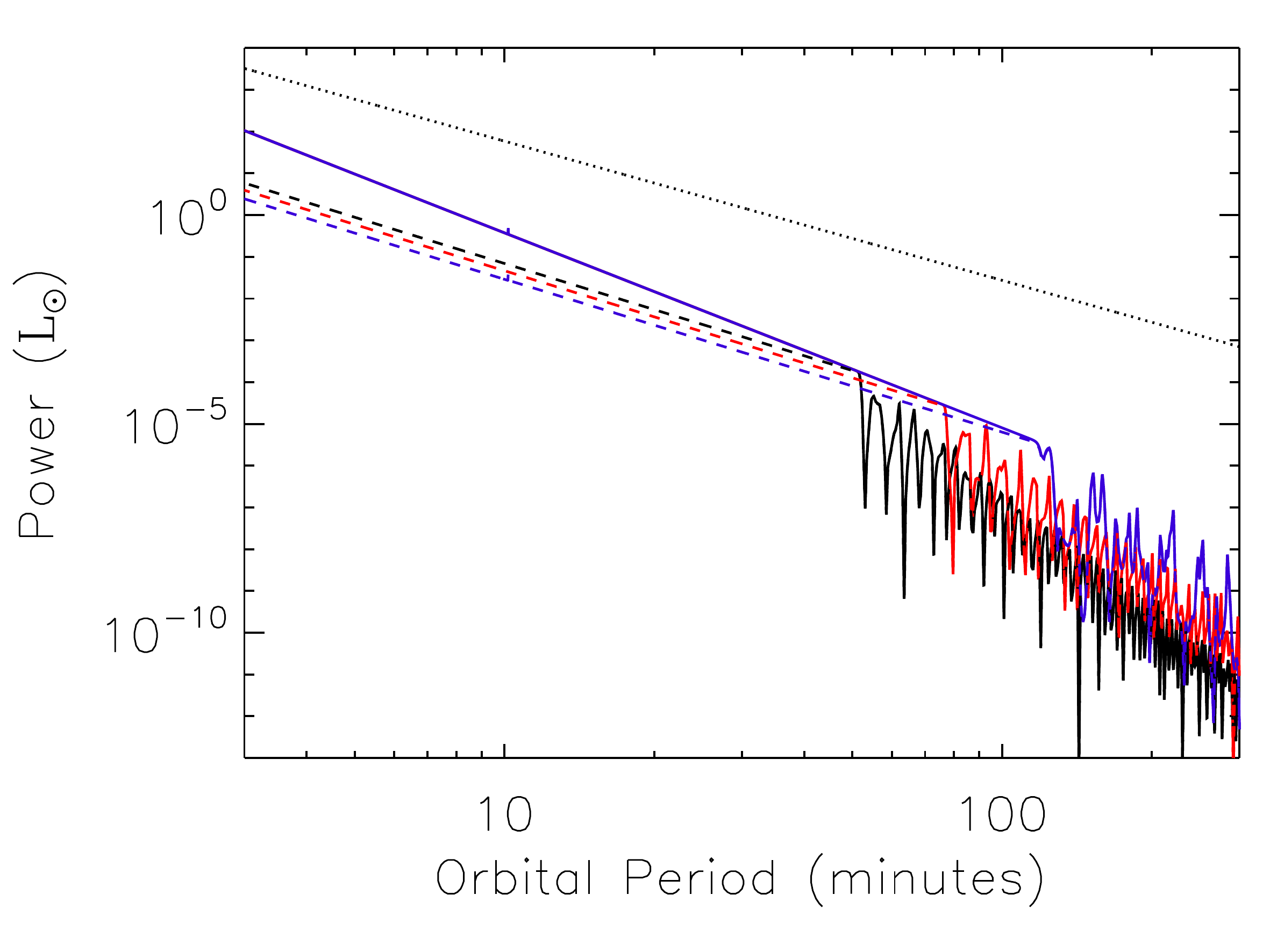}
\caption{\label{energy} The tidal energy dissipation rate $\dot{E}_{\rm tide}$ (solid lines) and the tidal heating rate $\dot{E}_{\rm heat}$ (dashed lines) as a function of orbital period. The black, red, and blue lines correspond to our WD models with $T_{\rm eff}=10800$K, $T=6000$K, and $T=3300$K, respectively. Note that at small orbital periods, the $\dot{E}_{\rm tide}$ curves overlap for different WD models. The dotted line is the energy dissipation rate due to gravitational waves, $\dot{E}_{\rm GW}$. In these evolutions, $M'=M$ and the WDs initially have $\Omega_s=0$.}
\end{centering}
\end{figure*}

Figure \ref{energy} shows the tidal energy transfer rate (from the orbit to the WD)
$\dot E_{\rm tide}=T_0\Omega F(\omega)$. 
For $\Omega\lo \Omega_c$, $\omega\simeq 2\Omega$ (assuming 
$\Omega_s\ll\Omega$), we see that 
$\dot E_{\rm tide}$ depends on $\Omega$ in a rather erratic manner.
However, when $\Omega\go\Omega_c$, efficient tidal synchronization ensures
$\dot\Omega\simeq\dot\Omega_s$, or $3\Omega/2t_{\rm GW}\simeq
T_0F(\omega)/I$, and thus $\dot E_{\rm tide}$ simplifies to
\be
\label{eeq}
\dot{E}_{\textrm{tide}} \simeq \frac{3 I\Omega^2}{2t_{\rm GW}} 
\qquad \big({\rm for} \ \Omega \go \Omega_c \big).
\ee
Since $\dot E_{\rm tide}/\dot E_{\rm GW}\simeq 3I/(\mu a^2)\ll 1$,
the orbital decay is dominated by gravitational radiation.
Nevertheless, the orbital phase evolution is affected by the tidal energy transfer, and such a phase shift can be measurable for short period binaries such as the recently discovered 12 minute system SDSS J0651 (Brown et al. 2011; see Section 9).  Also, low-frequency ($10^{-4}-10^{-1}$ Hz) gravitational waveforms emitted by the binary, detectable by LISA, will deviate significantly from the
point-mass binary prediction. This is in contrast to the case of
neutron star binaries (NS/NS or NS/BH) studied previously 
(Reisenegger \& Goldreich 1994; Lai 1994; Shibata 1994; Ho \& Lai 1999; 
Lai \& Wu 2006; Flanagan \& Racine 2007), 
where the resonant mode amplitude is normally too small to affect the
gravitational waveforms to be detected by ground-based gravitational
wave detectors such as LIGO and VIRGO, tidal effects
only become important near the NS binary merger (e.g.,
Lai et al. 1994; Hinderer et al.~2010).

The orbital cycle of a WD binary evolves according to  
\be
dN_{\rm orb}={\Omega\over 2\pi}{dE_{\rm orb}\over \dot E_{\rm orb}}.
\ee
Including tidal effects in $\dot E_{\rm orb}$, we find
\be
{dN_{\rm orb}\over d\ln \Omega}=
\left({dN_{\rm orb}\over d\ln \Omega}\right)_0\left(1+{\dot E_{\rm tide}
\over \dot E_{\rm GW}}\right)^{-1},
\ee
where 
\begin{align}
\left({dN_{\rm orb}\over d\ln \Omega}\right)_{\!0}&={\Omega t_{\rm GW}\over
3\pi}={5c^5\over 192\pi G^{5/3}\mu M_t^{2/3}(\pi f_{\rm GW})^{5/3}}\nonumber\\
&\!\!\!\!\!\!\!\!\!\!=2.3\!\times\!10^9\bigg(\frac{M_{\odot}^2}{MM'}\bigg)\bigg(\frac{M_t}
{2M_{\odot}}\bigg)^{\!\!1/3}\!\! \bigg(\!\frac{f_{\rm GW}}{0.01\,
\textrm{Hz}}\!\bigg)^{\!\! -5/3}
\end{align}
is the usual result when the tidal effect is neglected
($f_{\rm GW}=\Omega/\pi$ is the gravitational wave frequency).
Thus, even though $\dot E_{\rm tide}/\dot E_{\rm GW}\ll 1$, 
the number of ``missing cycles'' due to the tidal effect,
\be
\left({d\Delta N_{\rm orb}\over d\ln \Omega}\right)_{\rm tide}
\simeq -\left({dN_{\rm orb}\over d\ln \Omega}\right)_0
{\dot E_{\rm tide}\over \dot E_{\rm GW}},
\ee
can be significant. Since $E_{\rm tide}\propto I$, proper modelling and detection of the missing cycles would provide a measurement of the moment of inertia of the WD.

%%%%%%%%%%%%%
\subsection{Tidal Heating}

The tidal energy transfer $\dot{E}_\textrm{tide}$ does not correspond
to the energy dissipated as heat in the WD, because some of the energy
must be used to spin up the WD. Assuming rigid-body rotation, the tidal
heating rate is 
\be
\label{Eheat}
\dot{E}_\textrm{heat} = \dot{E}_\textrm{tide} \bigg(1 -
\frac{\Omega_s}{\Omega} \bigg).  
\ee
Figure \ref{energy} shows $\dot E_{\rm heat}$ for our three binary WD models.
At large binary seperations ($\Omega\lo\Omega_c$)
when $\Omega_s \ll \Omega$, virtually all of the tidal energy transfer 
to the WD is dissipated as heat. At smaller serparations, we 
have shown that the WD will retain a small degree of asynchronization.
Inserting equation (\ref{omsapprox}) into equation (\ref{Eheat}), 
we find
\begin{align}
\label{Eheat2}
\dot{E}_\textrm{heat} &\simeq \dot{E}_\textrm{tide}\bigg(\frac{\Omega_c}{\Omega}\bigg)^{16/15} \nonumber \\
&\simeq {3I\Omega^2\over 2t_{\rm GW}}\bigg(\frac{\Omega_c}{\Omega}\bigg)^{16/15} \qquad ({\rm for}~\Omega\go\Omega_c).
\end{align}
Thus, as the orbital frequency increases, a smaller fraction of the 
tidal energy is dissipated as heat. Using equation (\ref{omc}) for $\Omega_c$, 
we have 
\begin{align}
\dot E_{\rm heat}\simeq &(6.1\times 10^{36}\,{\rm erg~s}^{-1})
\kappa_{0.17}^{6/5}\hat f^{-1/5}M_1^{29/10}\nonumber \\
&\times(M_1')^{4/5}R_4^{-1/10} \left({\Omega\over 0.1~{\rm s}^{-1}}\right)^{18/5}.
\label{Eheat3}
\end{align}
Note that $\dot E_{\rm heat}$ is relatively insensitive to ${\hat f}$, 
so its precise value is not important. 
Thus, tidal heating of the WD can become significant
well before merger. For example, for our $T_{\rm eff}=10800$K WD model
(with $M=M'=0.6M_\odot$, $R=8970$~km and $\hat f\sim 20$), we find
$\dot E_{\rm heat}\sim 1.2\times 10^{32}$~erg/s at the orbital period
$P=10$~min, much larger than the ``intrinsic'' luminosity of the WD,
$4\pi R^2\sigma_{\rm SB}T_{\rm eff}^4=3.3\times 10^{30}$~erg/s.
Note that $\dot E_{\rm heat}$ is mainly deposited in the WD envelope, so 
an appreciable fraction of $\dot E_{\rm tide}$ may be radiated, and the WD
can become very bright prior to merger. The 12 minute binary SDSS J0651 (Brown et al. 2011) may be an example of such tidally heated WDs (see Section 9).

%%%%%%%%%%%%%%%%%%%%%%%%%%%%%%%%%%%%
\section{Discussion}

We have studied the tidal excitation of gravity waves in binary white
dwarfs (WDs) and computed the energy and angular momentum transfer
rates as a function of the orbital frequency for several WD models.
Such dynamical tides play the dominant role in spinning up the WD as
the binary decays due to gravitational radiation.  Our calculations
are based on the outgoing wave boundary condition, which implicitly
assumes that the tidally excited gravity waves are damped by nonlinear
effects or radiative diffusion as they propagate towards the WD
surface.  Unlike dynamical tides in early-type main-sequence stars,
where gravity waves are excited at the boundary between the convective
core and radiative envelope, the excitation of gravity waves in 
WDs is more complicated due to the various sharp features associated
with composition changes in the WD model.  We find that the
tidal energy transfer rate (from the orbit to the WD) 
$\dot E_{\rm tide}$ is a complex function of the tidal frequency
$\omega=2(\Omega-\Omega_s)$ (where $\Omega$ and $\Omega_s$ are the
orbital frequency and spin frequency, respectively; see Figures \ref{WD0090Eflux}-\ref{WD0160Eflux}),
and the local maxima of $\dot E_{\rm tide}$ scale approximately as
$\Omega^5\omega^5$. For most tidal frequencies considered, the gravity waves are
excited near the boundary between the carbon-oxygen core and the helium layer
(with the associated dip and sharp rise in the Brunt-V\"ais\"al\"a frequency profile).
We have constructed a semi-analytic model that captures 
the basic physics of gravity wave excitation 
and reveals that the complex behavior of 
$\dot E_{\rm tide}$ as a function of the tidal frequency
arises from the partial trapping of gravity waves in the quasi-resonance
cavity provided by the carbon-oxygen core.

We have also calculated the spin and orbital evolution of the WD
binary system including the effects of both gravitational radiation
and tidal dissipation. We find that above a critical orbital frequency
$\Omega_c$ [see equation (\ref{omc})], corresponding to an orbital period of
about an hour for our WD models, the dynamical tide BEGINS to
drive the WD spin $\Omega_s$ towards synchronous rotation, although a
small degree of asynchronization is maintained even at small
orbital periods: $\Omega-\Omega_s\simeq \Omega_c
(\Omega_c/\Omega)^{1/15}$ [see equation (\ref{omsapprox})].  Thus, numerical
simulations of WD binary mergers should use synchronized
configurations as their initial condition -- these may affect the
property of the merger product and possible supernova signatures. 

We also show that, although gravitational radiation always dominates over
tides in the decay of the binary orbit, tidal effects can nevertheless affect the orbital decay and
introduce significant phase error to the low-frequency gravitational
waveforms. Future detection of gravitational waves from WD binaries by
LISA may need to take these tidal effects into account and may lead to measurements of the WDs' moments of inertia.  Finally,
we have calculated the tidal heating rate of the WD as a function of
the orbital period. For $\Omega\go \Omega_c$, since the tidal
dissipation rate is largely controlled by the orbital decay rate due to
gravitational radiation, it is a smooth function of orbital period
[see equation (\ref{Eheat2})]. We show that well before mass transfer or
binary merger occurs, tidal dissipation in the WD envelope can be
much larger than the intrinsic luminosity of the star.
Thus, the WD envelope may be heated up significantly,
leading to brightening of the WD binary well before merger. We plan to study this issue in detail in a future paper.

The recently discovered 12 minute WD binary SDSS J0651 (Brown et al. 2011) can provide useful constraints for our theory. Applying equation (\ref{omc}) to this system, we find that the orbital period (12.75 minutes) is sufficiently short that both WDs are nearly (but not completely) synchronized with the orbit. Because of the orbital decay $\dot P$, the eclipse timing changes according to the relation 
\be
\Delta t=\dot{P} t^2/(2P), 
\ee
where $t$ is the observing time. Gravitational radiation gives rise to $\Delta t_{GW}=5.6{\rm s} \ (t/{1\rm yr})^2$.  Using equation (\ref{eeq}) to evaluate the orbital decay rate $\dot P_{\rm tide}$ due to tidal energy transfer, we find $\Delta t_{\rm tide}\simeq 0.28{\rm s} \ (t/{1\rm yr})^2$ (see also Benacquista 2011). Thus, the orbital decay due to tidal effects should be measured in the near future.
Also, our calculated heating rate, equation (\ref{Eheat3}), indicates the SDSS J0651 
WDs have suffered significant tidal heating, although to predict the luminosity change due to tidal heating requires careful study of the thermal structure of the WDs and knowledge of the location of tidal heating.
We note that Piro (2011) also considered some aspects of tidal effects in SDSS J0651, but his results were based on parameterized equilibrium tide theory.

This paper, together with paper I, represents only the first study of
the physics of dynamical tides in compact WD binaries, and more works are needed.
We have adopted several approximations that may limit the
applicability of our results. First, we have not included
the effects of rotation (e.g., the Coriolis force) in our wave equations. 
In addition to modifying the properties of gravity waves 
(they become generalized Hough waves), rotation also introduces 
inertial waves that can be excited once the WD spin frequency becomes 
comparable to the tidal frequency -- this may lead to more efficient
tidal energy transfer and synchronization. For example, if we parameterize
the spinup rate due to various mechanisms (including inertial waves)
by equation (87), the critical orbital frequency for the onset of synchronization
($\Omega_c$) is given by equation (\ref{omegac2}). For a stronger tidal torque
(larger $A$), $\Omega_c$ is smaller. However, the tidal heating rate
at $\Omega\go\Omega_c$ becomes [cf. equation (\ref{Eheat2})]
\be
\dot E_{\rm heat}\simeq {3I\Omega^2\over t_{\rm GW}}\left({\Omega_c
\over\Omega}\right)^{(3n+1)/(3n)}.
\ee
Thus, for stronger tidal torques, at a given orbital frequency ($\Omega \go\Omega_c)$, the tidal heating rate is reduced because the WD is closer to synchronization.

Second, we have assumed that the WD rotates as a rigid body.  
As the tidally-excited gravity waves deposit angular momentum 
in the outer layer of the WD, differential rotation will develop
if the different regions of the WD are not well coupled.
Thus it may be that the outer layer becomes synchronized 
with the companion while the core rotates
at a sub-synchronous rate, analogous to tidal synchronization
in early-type main-sequence stars (Goldreich \& Nicholson 1989).
Third, we have implicitly assumed that the outgoing gravity waves
are efficiently damped near the WD surface. This may not apply for
all WD models or all orbital frequencies. If partial wave reflection
occurs, tidal dissipation will be reduced compared to the results
presented in this paper except when the tidal frequency
matches the intrinsic frequency of a g-mode (cf.~Paper I).
More detailed studies on nonlinear wave damping (e.g., 
Barker \& Ogilvie 2010; Weinberg et al.~2011) and
radiative damping would be desirable.

Finally, we have only studied carbon-oxygen WDs in this paper.
Our calculations have shown that the strength of dynamical tides
depends sensitively on the detailed internal structure of the WD.
Recent observations (see references in Section 1)
have revealed many compact WD binaries 
that contain at least one low-mass helium-core WD. 
The temperatures of these helium-core WDs tend to be high ($T_{\rm eff}\go 10^4$K).
These observations warrant investigation of tidal effects in hot, 
helium-core WDs, which have significantly different internal structures from
the cool, carbon-oxygen WDs considered in this paper.

\section*{Acknowledgments}

We thank Gilles Fontaine (University of Montreal) for providing the white
dwarf models used in this paper and for valuable advice on these
models.  DL thanks Lars Bildsten and Gordon Ogilvie for
useful discussions, and acknowledges the hospitality (Spring 2010) 
of the Kavli Institute for Theoretical Physics at UCSB (funded by 
the NSF through Grant PHY05-51164) where part of the work was carried out.
This work has been supported in part by NSF grant AST-1008245.

\def\apj{{Astrophys. J.}}
\def\apjs{{Astrophys. J. Supp.}}
\def\mnras{{Mon. Not. R. Astr. Soc.}}
\def\prl{{Phys. Rev. Lett.}}
\def\prd{{Phys. Rev. D}}
\def\apjl{{Astrophys. J. Let.}}
\def\pasp{{Publ. Astr. Soc. Pacific}}
\def\aapr{{Astr. Astr. Rev.}}

%%%%%%%%%%%%%%%%%%%%%% 

\appendix
\section{Calculation with Massive Star Model}
\label{massive}

\begin{figure*}
\begin{centering}
\includegraphics[scale=.6]{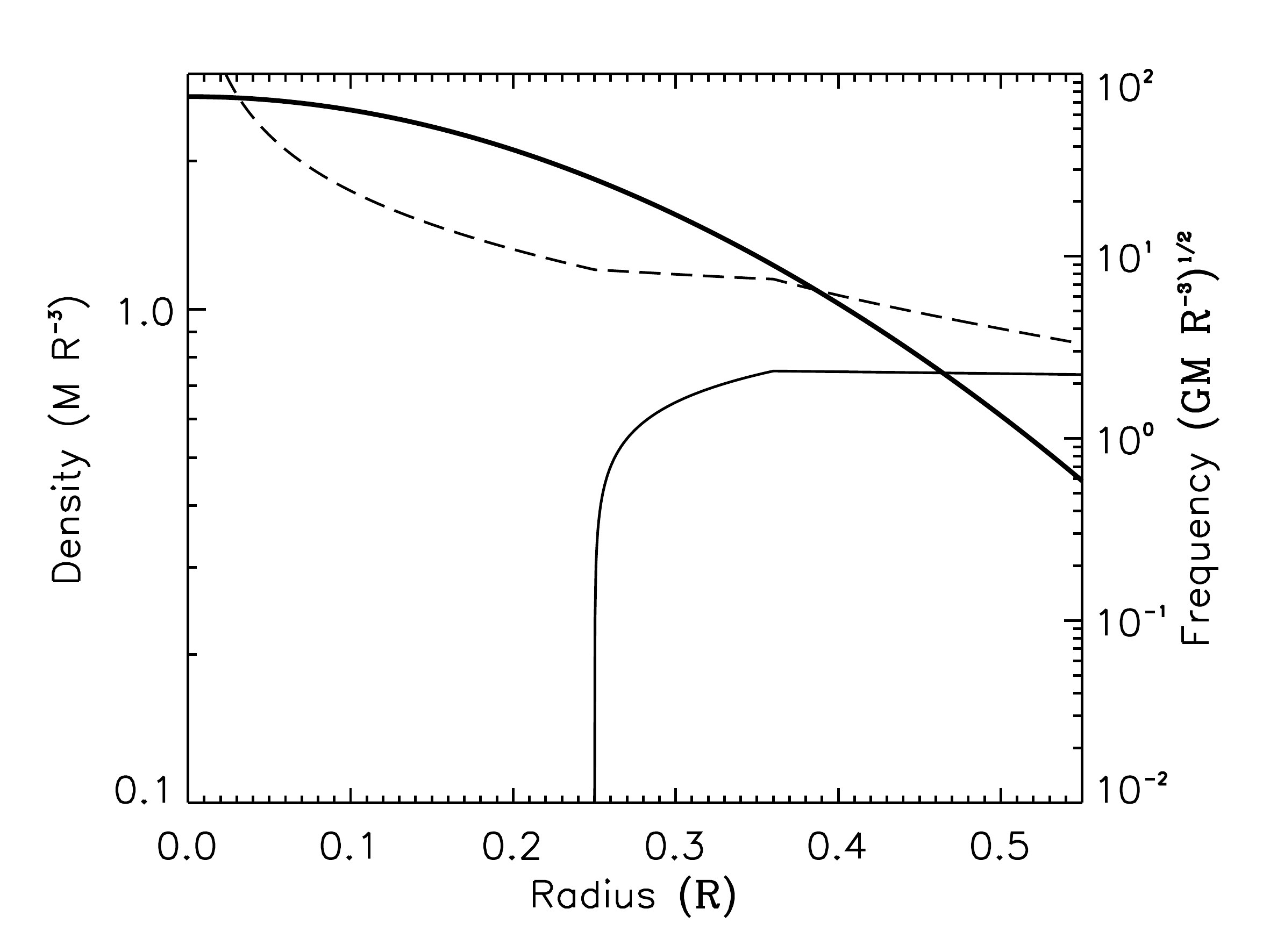}
\caption{\label{massivestar} The square of the Br\"unt-Vais\"al\"a (thin solid line) and Lamb (dashed line) frequencies (for $l=2$), in units of $GM/R^3$, as a function of the normalized radius in a simple massive star model. Also plotted is the stellar density profile (thick solid line) in units of $M/R^3$. The model has an inner convection zone extending to $r=0.25R$. The stellar properties are only plotted out to $r= 0.6R$, where an outgoing wave boundary condition is adopted in our calculation of the tidal excitation.}
\end{centering}
\end{figure*}

To test the accuracy of our numerical calculations and especially the importance of self-consistency in real stellar models, we compute the tidal response of several toy models. The first toy model we employ is shown in Figure \ref{massivestar} and is meant to mimic a massive early-type star. The model contains an inner convection zone surrounded by a thick radiative envelope. The convection zone extends to $r=0.25 R$, beyond which the value of $N^2$ rises linearly to $N^2 \approx 8 GM/R^3$. Dynamical tides in such massive stars have been studied by Zahn (1975, 1977) and Goldreich \& Nicholson (1989), who showed that the dominant effect arises from the gravity waves launched at the core-envelope boundary, which then propagate outwards and eventually dissipate near the stellar surface. Zahn (1975) derived an analytic solution for the wave amplitude and the corresponding tidal torque. Although our model does not contain some of the details exhibited by realistic massive star models, it does capture the most important features. We can compare our result with Zahn's to calibrate our numerical method and to assess the degree of self-consistency required to produce reliable results for the tidal torque.

\begin{figure*}
\begin{centering}
\includegraphics[scale=.7]{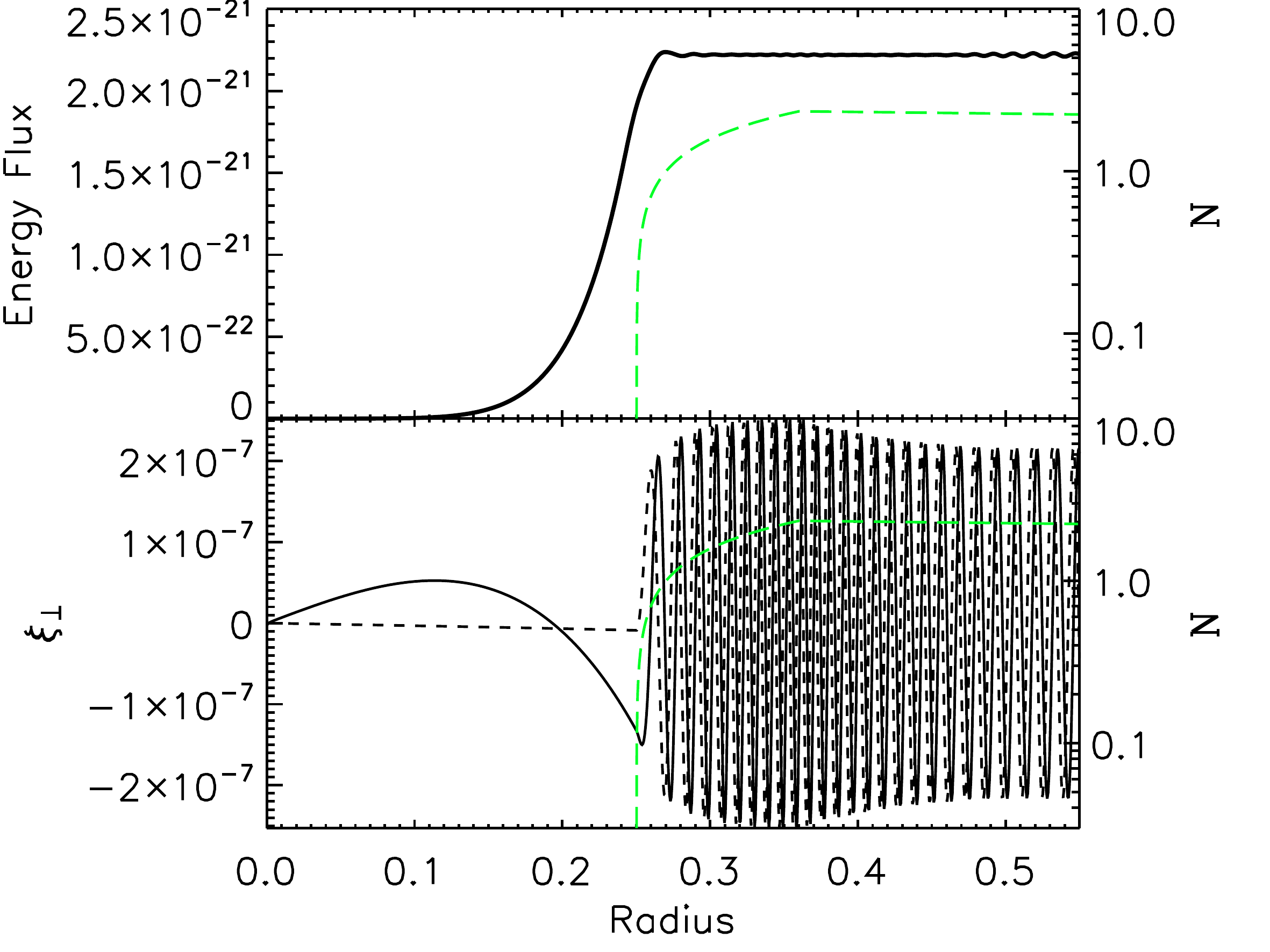}
\caption{\label{RadialEfluxMassiveStar} Dynamical tide in a massive star (based on the toy model depicted in Figure \ref{massivestar}) driven by a companion of mass $M'=M$, with the tidal frequency $\omega=2.3\times10^{-2}$. Top: The energy flux (dark solid line) $\dot{E} = \Omega \dot{J}_z$ as a function of radius, with $\dot{J}_z$ calculated from equation (\ref{Lz5}). All values are plotted in units of $G=M=R=1$. Bottom: The real part of $\xi_\perp^{\rm{dyn}}$ (dark solid line) and imaginary part of $\xi_\perp^{\rm{dyn}}$ (dashed line) as a function of stellar radius. The value of $N$ has been plotted in green (light solid line) in both panels. In this model, the energy flux rises to its final value just outside of the convective zone, showing that the wave is excited at this location.}
\end{centering}
\end{figure*}

\begin{figure*}
\begin{centering}
\includegraphics[scale=.6]{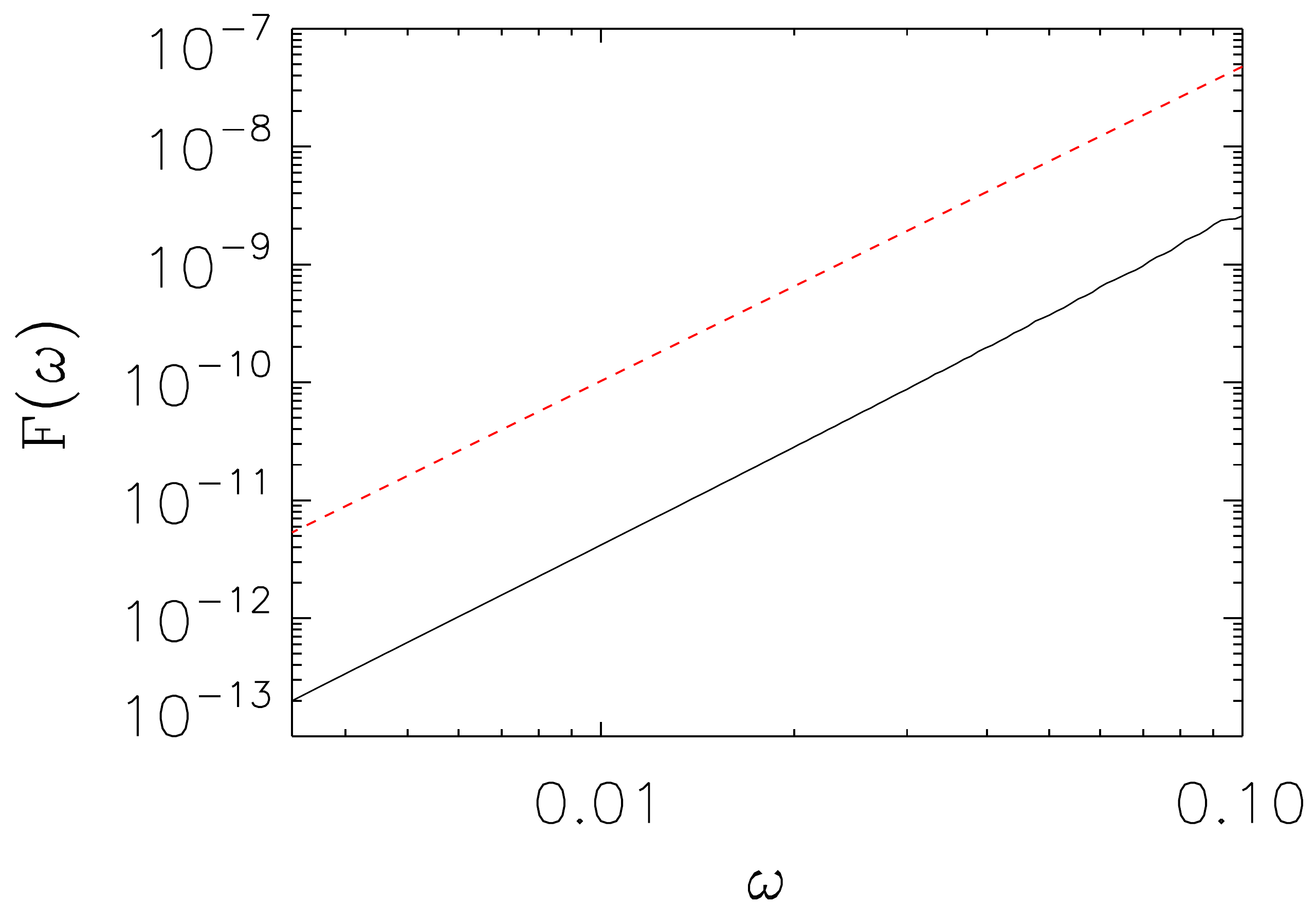}
\caption{\label{EfluxMassiveStar} The dimensionless tidal torque $F(\omega) = \dot{J}_z/T_o$ [see equation (\ref{edotf})] carried by the outgoing gravity wave as a function of the tidal frequency $\omega$ (solid line). The analytical estimate from equation (\ref{massivescale}) is also plotted (dashed line). The frequency is in units of $G=M=R=1$. The small wiggles at high frequencies are likely due to the slight inaccuracy of our implementation of the outer boundary condition due to the neglected terms which become non-negligible at higher tidal frequencies.}
\end{centering}
\end{figure*}

Figure \ref{RadialEfluxMassiveStar} shows an example of our numerical results for the dynamical tides generated in a massive star by a companion, for a given tidal frequency $\omega = 2 \Omega= 2.3\times10^{-2}$ (in units where $G=M=R$). We see that gravity waves are excited at the base of the radiative zone where $N^2$ begins to rise above zero. A net energy flux $\dot{E} = \Omega \dot{J}_z = \Omega(GM'^2R^5/a^6)F(\omega)$ flows outwards toward the stellar surface. Figure \ref{EfluxMassiveStar} shows our numerical result of the dimensionless function $F(\omega) \equiv \dot{J}_z/T_o$ [see equation (\ref{edotf})], evaluated at the outer boundary, as a function of the tidal frequency $\omega$. The result can be fitted by $F(\omega) \propto \omega^{8/3}$, in agreement with the scaling found by Zahn (1975).

The power-law scaling of the energy flux can be derived using the method of Goldreich \& Nicholson (1989). Assume $|\xi_\perp^{\rm dyn}| \approx \xi_\perp^{\rm{eq}}$ at $r=r_{c+}$, which is located one wavelength above the convective boundary ($r=r_c$).  From the dispersion relation (\ref{k2approx}), we find that the Br\"unt-Vais\"al\"a frequency at $r_{c+}$ is given by (for $l=2$) 
\be
\label{}
N(r_{c+}) \approx \bigg(\frac{d N^2}{dr} r_c\bigg)^{1/3} \omega^{1/3}.
\ee
Using $\xi_\perp^{\rm{eq}} \simeq -\big[1/(6r)\big] \big(Ur^2/g\big)' \simeq -U/(2g)$, we evaluate equation (\ref{Lscale}) to find
\be
\label{massivescale}
\dot{E} \approx \frac{3\pi\sqrt{6}}{10} \bigg(\frac{M'}{M_t}\bigg)^2 \frac{\rho r^7 \Omega^5 \omega^{8/3}}{g^2 (dN^2/d\ln r)^{1/3}}.
\ee
where $M_t=M+M'$, and all the quantities ($\rho$, $r$, $g$, and $dN^2/dr$ are evaluated at $r=r_{c+} \simeq r_c$). The scaling of this estimate nearly agrees Goldreich \& Nicholson (1989), who obtained $\dot{E}_r \propto \Omega^4 \omega^{11/3}$, where $\dot{E}_r$ is the energy flux carried by outgoing gravity waves in the rotating frame of the star, not the total energy transfer rate from the orbit. These two energy transfer rates are related by $\dot{E} = \Omega \dot{J_z} = 2\Omega \dot{E}_r/\omega$. Goldreich \& Nicholson (1989) estimates $dN^2/dr \approx g/H \approx g/r$, and with $g \simeq 4\pi G \bar{\rho} r/3$ ($\bar{\rho}$ is the mean density interior to $r_c$), equation (\ref{massivescale}) becomes
\be
\label{massivescale2}
\dot{E} \approx 0.08 \bigg(\frac{M'}{M_t}\bigg)^2 \frac{\rho r^5 \Omega^5 \omega^{8/3}}{(G \bar{\rho})^{7/3}}.
\ee
The value of $F(\omega)$ based on equation (\ref{massivescale}) is plotted in Figure \ref{EfluxMassiveStar}. Compared to our numerical results, we see that equation (\ref{massivescale}) overestimates $F(\omega)$ by an order of magnitude [by contrast, equation (\ref{massivescale2}) would overestimate $F(\omega)$ by more, since for our toy stellar model $dN^2/d \ln r \gg g/r$]. From our numerical results, we find that the dynamical part of the tide only reaches an amplitude of $\xi_\perp^{\rm dyn} \approx \xi_\perp^{\rm eq}/4$. If we had used this wave amplitude in our estimate, equation (\ref{massivescale}) would be a factor of 16 smaller and would provide an accurate approximation to $F(\omega)$ at all frequencies considered.

\end{document}